\documentclass[onecolumn]{IEEEtran}
\usepackage{graphics,color,subfigure,latexsym,amssymb,amsmath,url}
\usepackage{bbm}
\usepackage[pdftex]{graphicx} 

\newcommand{\eqdef} {\mathrel{\mathop:}=}
\newtheorem{definition}{Definition}
\newtheorem{remark}{Remark}
\newtheorem{theorem}{Theorem}

\newtheorem{lemma}{Lemma}

\newtheorem{corollary}{Corollary}

\newcommand{\be}{\begin{equation}}
\newcommand{\ee}{\end{equation}}
\newcommand{\ben}{\begin{enumerate}}
\newcommand{\een}{\end{enumerate}}
\newcommand{\bea}{\begin{eqnarray}}
\newcommand{\eea}{\end{eqnarray}}
\newcommand{\bean}{\begin{eqnarray*}}
\newcommand{\eean}{\end{eqnarray*}}

\def\squarebox#1{\hbox to #1{\hfill\vbox to #1{\vfill}}}
\def\eor{\hskip 3pt $\Diamond$}

\newcommand{\cX}{{\cal X}}
\newcommand{\cY}{{\cal Y}}

\newcommand{\cP}{{\cal P}}
\newcommand{\cD}{{\cal D}}
\newcommand{\cB}{{\cal B}}
\newcommand{\cS}{{\cal S}}

\def\b0{\mathbf{0}}
\def\bX{\mathbf{X}}
\def\bZ{\mathbf{Z}}
\def\bW{\mathbf{W}}

\def\bT{\mathbf{T}}
\def\bt{\mathbf{t}}

\def\bY{\mathbf{Y}}

\def\bZ{\mathbf{Z}}
\def\bA{\mathbf{A}}

\def\bW{\mathbf{W}}

\def\bbR{\mathbb{R}}
\def\bbZ{\mathbb{Z}}
\def\bx{\mathbf{x}}
\def\by{\mathbf{y}}
\def\bz{\mathbf{z}}

\def\bb{\mathbf{b}}
\def\bc{\mathbf{c}}
\def\bd{\mathbf{d}}

\def\bv{\mathbf{v}}

\def\bS{\mathbf{\Sigma}}

\def\mE{\textrm{E}}
\def\me{\textrm{e}}
\def\mI{\textrm{I}}
\def\mD{\textrm{D}}
\def\mP{\textrm{P}}
\def\me{\textrm{e}}
\def\mV{\textrm{Var}}
\def\mSP{\textrm{SP}}
\def\mcr{\textrm{cr}}
\def\mr{\textrm{r}}
\def\mF{\textrm{F}}
\def\mo{\textrm{o}}
\def\b1{\mathbbm{1}}
\def\W1{W_{1^{-},P,R}}

\def\d{\partial}
\def\eor{\hskip 3pt $\Diamond$}
\def\eot{\hskip 3pt $\blacklozenge$}
\def\eod{\hskip 3pt $\lozenge$}

\begin{document}

\title{Refinement of the random coding bound}
\author{Y\"{u}cel Altu\u{g}~\IEEEmembership{Member }and Aaron B.~Wagner~\IEEEmembership{Senior Member}
\thanks{
The material in this paper was presented in part at the 50th Annual Allerton Conference on Communication, Control, and Computing and 2012 International Zurich Seminar on Communications. 

The authors are with the School of Electrical and Computer Engineering, Cornell University, 
Ithaca, NY 14853.
E-mail: {\em ya68@cornell.edu, wagner@ece.cornell.edu}.}}

\maketitle
\begin{abstract}
An improved pre-factor for the random coding bound is proved. Specifically, for channels with critical rate not equal to capacity, if a regularity condition is satisfied (resp. not satisfied), then for any $\epsilon >0$ a pre-factor of $O(N^{-\frac{1}{2}\left( 1 - \epsilon + \bar{\rho}^\ast_R \right)})$ (resp. $O(N^{-\frac{1}{2}})$) is achievable for rates above the critical rate, where $N$ and $R$ is the blocklength and rate, respectively. The extra term $\bar{\rho}^\ast_R$ is related to the slope of the random coding exponent. Further, the relation of these bounds with the authors' recent refinement of the sphere-packing bound, as well as the pre-factor for the random coding bound below the critical rate, is discussed. 
\end{abstract}

\section{Introduction}
\label{sec:intro}
Deriving precise bounds on the optimal error probability of block codes over a discrete memoryless channel (DMC) is a long-established topic in information theory (see, e.g., \cite{feinstein54}--\cite{csiszar-korner81} and references therein). Traditionally, the focus of this effort has been on the asymptotic regime in which the rate is held fixed below capacity and the blocklength tends to infinity. The error probability decays exponentially in this regime, and the goal has been to determine the best possible exponent, called the \emph{reliability function} of the channel. 

Classical results \cite{fano61}--\cite{haroutunian68} provide upper and lower bounds on the error probability in this regime. These bounds both decay exponentially fast, and in fact their exponents coincide at high data rates. Unfortunately, however,  there is a sizeable gap between the sub-exponential factors in these bounds. Specifically, until recently, the best known\footnote{We consider the bounds that are valid for any DMC. For some specific DMCs, improved bounds are available~\cite{elias55}.} sub-exponential factors of the random coding upper bound and sphere-packing bound lower bound were $O(1)$ and $\Omega(N^{-|\cX||\cY|})$, due to Fano~\cite{fano61} and Haroutunian~\cite{haroutunian68}, respectively, where $\cX$ and $\cY$ are input and output alphabets of the channel. 

At rates close to capacity, the sub-exponential pre-factor can potentially have a large effect on the error probability bound, since the exponent is known to vanish as the rate approaches capacity. Of course, rates close to capacity are also of greatest interest from a practical standpoint.
Indeed, the invention of practical capacity-achieving codes (e.g., \cite{turbo93}--\cite{arikan2009}) have made the derivation of accurate error probability bounds for rates close to capacity an important practical concern. Due to the large gap between the exponential pre-factors in the upper and lower bounds mentioned above, the classical error exponents bounds do not provide accurate design guidelines in the vicinity of capacity~\cite[Section~V]{polyanskiy10}.

In this work, our goal is to improve the achievable pre-factors in front of the exponentially decaying term of the random coding bound (complementing recent analogous work on the sphere-packing bound~\cite{altug12-e}). To do so, we revisit the random coding arguments of Fano~\cite{fano61} and refine them to provide an improved pre-factor. Our variation distinguishes between two types of channels that collectively exhaust all DMCs. Our main findings are
\begin{enumerate}
\item 
\label{item:result1} If a DMC with the critical rate is not equal to capacity\footnote{For the definition of the critical rate, see \cite[pg.~160]{gallager68}.} satisfies a certain condition, then for rates between the critical rate and capacity, there exists an $(N,R)$ code with maximal error probability smaller than 
\begin{equation}
\frac{K_1 e^{-N \mE_\mr(R)}}{N^{\frac{1}{2}\left( 1 -\epsilon + \bar{\rho}_R^\ast \right)}},
\label{eq:intro1}
\end{equation}
for any $\epsilon >0$, where $K_1$ is a positive constant that depends on the channel, rate and $\epsilon$, and $\bar{\rho}_R^\ast$ is related to the subdifferential of the random coding exponent $\mE_\mr(R)$. Further, if the channel is positive, then $\bar{\rho}_R^\ast$ is the \emph{left derivative} of $\mE_\mr(R)$ and one can drop the $\epsilon$ in \eqref{eq:intro1}. 

\item 
\label{item:result2} If a DMC with the critical rate is not equal to capacity does not satisfy the aforementioned condition\footnote{A canonical example of this type of channels is binary erasure channel.}, then for rates between capacity and the critical rate, there exists an $(N,R)$ code with maximal error probability smaller than $$\frac{K_2 e^{-N \mE_\mr(R)}}{\sqrt{N}},$$ where $K_2$ is a positive constant that depends on the channel and rate. 
\end{enumerate}

In a forthcoming paper~\cite{altug13}, we shall show that for 
symmetric channels,
the order of the pre-factor in both bounds is tight in the sense
that one can prove lower bounds of the same form that hold for
all codes with rate $R$.  For asymmetric
channels, it is worth noting that 
the upper bound in item \ref{item:result1}) is very close to a lower bound
recently established by the 
authors for arbitrary constant composition codes~\cite{altug12-e} 
(see (\ref{eq:sp}) to follow). 

The upper bounds in items
\ref{item:result1}) and \ref{item:result2}) are established by first 
proving an upper bound, with an exponent of $\mE_\mr(R,Q)$, on
the error probability of a random code whose codewords are drawn
i.i.d.\ according to some distribution $Q$.
We also determine the exact order 
of the pre-factor for random coding below the critical rate,
correcting a small error in the literature.
After the conference versions~\cite{altug12-b, altug-allerton12} 
of this work appeared, Scarlett 
\emph{et al.}~\cite{scarlett13, scarlett13-a} 
generalized the
main results in several directions. They also provided a shorter proof of the 
results as stated here. Although longer, we believe our original 
proof is more amenable to analysis of non-i.i.d.\ code ensembles,
as described in Remark~\ref{item:LDPC}. It also provides some 
intuition as to why the $\bar{\rho}_R^\ast$ term appears in the
pre-factor in case~\ref{item:result1}).


\section{Notation, Definitions and Statement of the Result}
\label{sec:notation-definitions}
\subsection{Notation}
\label{ssec:notation}
Boldface letters denote vectors, boldface letters with subscripts denote individual components of vectors. Furthermore, capital letters represent random variables and lowercase letters denote individual realizations of the corresponding random variable. Throughout the paper, all logarithms are base-$e$. For a finite set $\cX$, $\cP(\cX)$ denotes the set of all probability measures on $\cX$. Similarly, for two finite sets $\cX$ and $\cY$, $\cP(\cY|\cX)$ denotes the set of all stochastic matrices from $\cX$ to $\cY$. 
$\bbR$, $\bbR^{+}$ and $\bbR_{+}$ denote the set of real, positive real and non-negative real numbers, respectively. $\bbZ^{+}$ denotes the set of positive integers. We follow the notation of the book of Csisz\'ar-K\"orner \cite{csiszar-korner81} for standard information theoretic quantities.  
\subsection{Definitions}
\label{ssec:defn}
Throughout the paper, let $W$ be a DMC from $\cX$ to $\cY$. For any $Q \in \cP(\cX)$, 
\begin{equation}
\mE_{\mr}(R,Q) \eqdef \max_{0 \leq \rho \leq 1}\left\{ -\rho R + \mE_{\mo}(\rho, Q) \right\}, 
\label{eq:ErQ}
\end{equation} 
where
\begin{equation}
\mE_{\mo}(\rho, Q) \eqdef -\log \sum_{y \in \cY}\left[ \sum_{x \in \cX}Q(x)W(y|x)^{1/(1+\rho)}\right]^{(1+\rho)}. 
\label{eq:Eo}
\end{equation}
The \emph{random coding exponent} is defined as 
\begin{equation}
\mE_{\mr}(R) \eqdef \max_{Q \in \cP(\cX)} \mE_{\mr}(R,Q). 
\label{eq:Er}
\end{equation}

For any $W \in \cP(\cY|\cX)$, $Q \in \cP(\cX)$, $N \in \bbZ^+$ and $R \in \bbR_+$ the \emph{ensemble average error probability} of an $(N,R)$ random code with codewords generated by using $Q$ along with a maximum likelihood decoder\footnote{We assume that ties always lead to an error. However, this pessimistic assumption increases the error probability by at most a factor of $2$.} is denoted by $\bar{\mP}_\me(Q,N,R)$. For any $(N,R)$ code $(f, \varphi)$, $\textrm{P}_{\textrm{e}}(f, \varphi)$ denotes the maximal error probability of the code.

Further, define
\begin{align}
\cS_{Q} & \eqdef \left\{ (x,y) \in \cX \times \cY : Q(x)W(y|x) >0\right\}, \label{eq:SQ}\\
\tilde{\cS}_{Q} & \eqdef \left\{ (x,y,z) \in \cX \times \cY \times \cX : Q(x)W(y|x)Q(z)W(y|z) > 0 \right\}, \label{eq:tSQ}\\
\cX_y & \eqdef \left\{ x \in \cX : W(y|x) > 0\right\}. \label{eq:Xy}
\end{align}

Given a $(Q,W) \in \cP(\cX) \times \cP(\cY|\cX)$ pair, the following definition plays a crucial role in our analysis. 

\begin{definition}[Singularity]
\begin{itemize}
\item[(i)] A $(Q,W) \in \cP(\cX) \times \cP(\cY|\cX)$ pair is called \emph{singular} if 
\begin{equation}
W(y|x) = W(y|z), \, \forall \, (x,y,z) \in \tilde{\cS}_{Q}.
\label{eq:singular-defn}
\end{equation}
Otherwise, it is called \emph{nonsingular}. The set of all nonsingular (resp. singular) $(Q,W)$ pairs is denoted by $\cP_{\textnormal{ns}}$ (resp. $\cP_{\textnormal{s}}$).
\item[(ii)] A channel $W$ is called \emph{nonsingular at rate $R$} provided that there exists $Q \in \cP(\cX)$ with $\mE_{\mr}(R,Q) = \mE_{\mr}(R)$ such that  $(Q,W) \in \cP_{\textnormal{ns}}$. Similarly, a channel is called \emph{singular at rate $R$} if for all $Q \in \cP(\cX)$ with $\mE_{\mr}(R,Q) = \mE_{\mr}(R)$, $(Q,W) \in \cP_{\textnormal{s}}$. \eod
\end{itemize}
\label{def:singular}
\end{definition}

\begin{remark} Consider any $(Q,W) \in \cP(\cX) \times \cP(\cY|\cX)$ pair. 
\begin{itemize}
\item[(i)] Definition~\ref{def:singular} can be viewed as a condition that ensures that when a random code with distribution $Q$ is used for transmission through channel $W$, the optimal decoding algorithm, given the channel output, simply finds a ``feasible'' codeword. Indeed, in such a situation, all codewords with
nonzero posterior probability given the channel output have the same
posterior probability.

\item[(ii)] In his investigation of the zero undetected error capacity\footnote{For the definition of zero undetected error capacity, see \cite[pg.~42]{telatar92}.} of discrete memoryless channels, Telatar uses a property similar to Definition~\ref{def:singular}. In particular, he proves that the zero undetected error capacity is equal to (Shannon) capacity for  ``channels for which the non-zero values of $W(y|x)$ depend only on $y$'' \cite[pg.~51]{telatar92}. 
In our terminology, this is the set of channels $W$ for which
$(Q,W)$ is singular for all $Q$ or, equivalently, $(Q,W)$ is
singular when $Q$ is the uniform distribution.

\item[(iii)] Singularity also plays a significant role in the third-order term of the normal approximation for a DMC, as demonstrated in \cite{altug13-a}.

\item[(iv)] For an explanation of why we use the term \emph{singular}, see Remark~\ref{rem:rem-thrm1-pf-1}. \eor
\end{itemize}
\label{rem:rem1}
\end{remark}

Given $W \in \cP(\cY|\cX)$ with $R_{\mcr}<C$, and $R \in (R_{\mcr},C)$ such that $W$ is nonsingular at rate $R$, we define\footnote{Differentiability of $\mE_{\mr}(\cdot, Q)$ is proved in Lemma~\ref{lem:prelim} to follow.}
\begin{equation}
\bar{\rho}^{\ast}_R \eqdef \sup_{Q : \mE_{\mr}(R,Q) = \mE_{\mr}(R) \textrm{ and }(Q,W) \in \cP_{\textrm{ns}}} -\left.\frac{\partial \mE_{\mr}(a,Q)}{\partial a}\right|_{a =R}.
\label{eq:rho-star}
\end{equation}

\subsection{Results}
\label{ssec:result}
\begin{theorem}
Let $W \in \cP(\cY|\cX)$ be arbitrary with $R_{\mcr} < C$. 
\begin{itemize}
\item[(i)] If $Q \in \cP(\cX)$ and $R \in \bbR_+$ are such that the pair $(Q,W)$ is singular and\footnote{$R_\mcr(Q) \eqdef \left. \frac{\partial \mE_\mo(\rho, Q)}{\partial \rho}\right|_{\rho =1}$ (e.g., \cite[pg.~142]{gallager68}).} $R_\mcr(Q) < R < \mI(Q;W)$, then there exists $K_1 \in \bbR^+$ that depends on $W,R$ and $Q$ such that
\begin{equation}
\bar{\mP}_{\me}(Q,N,R) \leq \frac{K_1}{\sqrt{N}}e^{-N \mE_{\mr}(R,Q)},
\label{eq:thrm1.ii} 
\end{equation}
for all $N \in \bbZ^+$. Further, there exists an $(N,R)$ code $(f, \varphi)$ and $\tilde{K}_1 \in \bbR^+$ that depends on $W,R$ and $Q$ such that 
\begin{equation}
\mP_\me(f, \varphi) \leq \frac{\tilde{K}_1}{\sqrt{N}}e^{-N \mE_\mr(R,Q)},
\label{eq:thrm1.ii-1}
\end{equation}
for all $N \in \bbZ^+$. 

\item[(ii)] If $Q \in \cP(\cX)$ and $R \in \bbR_+$ are such that the pair $(Q,W)$ is nonsingular and $R_\mcr(Q) < R< \mI(Q;W)$, then there exists $K_2 \in \bbR^+$ that depends on $W,R$ and $Q$ such that
\begin{equation}
\bar{\mP}_{\me}(Q,N,R) \leq \frac{K_2}{N^{0.5(1+\rho^\ast_R(Q))}}e^{-N \mE_{\mr}(R,Q)}, 
\label{eq:thrm1.i}
\end{equation}
for all $N \in \bbZ^+$ where $\rho^\ast_R(Q) \eqdef - \left.\frac{\partial \mE_{\mr}(a, Q)}{\partial a}\right|_{a = R}$. Further, there exists an $(N,R)$ code $(f, \varphi)$ and $\tilde{K}_2 \in \bbR^+$ that depends on $W,R$ and $Q$ such that 
\begin{equation}
\mP_\me(f, \varphi) \leq \frac{\tilde{K}_2}{N^{0.5(1+\rho^\ast_R(Q))}}e^{-N \mE_{\mr}(R,Q)},
\label{eq:thrm1.i-1}
\end{equation}
for all $N \in \bbZ^+$. \eot
\end{itemize}
\label{thrm:thrm1}
\end{theorem}
\begin{IEEEproof}
Theorem~\ref{thrm:thrm1} is proved in Section~\ref{sec:proof-thrm1}. 
\end{IEEEproof}

Theorem~\ref{thrm:thrm1} immediately implies the following.
\begin{corollary}
Let $W \in \cP(\cY|\cX)$ be arbitrary with $R_\mcr < C$ and $R \in (R_\mcr, C)$. 
\begin{itemize}
\item[(i)] If $W$ is singular at rate $R$, then there exists an $(N,R)$ code $(f, \varphi)$ and $K_3 \in \bbR^+$ that depends on $R$ and $W$ such that 
\begin{equation}
\mP_\me(f, \varphi) \leq \frac{K_3}{\sqrt{N}}e^{-N \mE_{\mr}(R)},
\label{eq:cor1-i}
\end{equation}
for all $N \in \bbZ^+$. 
\item[(ii)] If $W$ is nonsingular at rate $R$, then for any $\epsilon >0$, there exists an $(N,R)$ code $(f, \varphi)$ and $K_4 \in\bbR^+$ that depends on $R,W$ and $\epsilon$ such that 
\begin{equation}
\mP_{\me}(f, \varphi) \leq \frac{K_4}{N^{0.5(1+\bar{\rho}^\ast_R - \epsilon)}}e^{-N \mE_\mr(R)}, 
\label{eq:cor1-ii}
\end{equation}
for all $N \in \bbZ^+$. \eot
\end{itemize}
\label{cor:RC-cor1}
\end{corollary}

One can omit the $\epsilon$ in the exponent in~(\ref{eq:cor1-ii}) if  the supremum in (\ref{eq:rho-star}) is achieved. The next result shows that, for the most channels, something even stronger is true.

\begin{theorem}
Let $W \in \cP(\cY|\cX)$ be arbitrary with $R_{\mcr} < C$ and $R \in (R_\mcr, C)$. 
\begin{itemize}
\item[(i)] The subdifferential of $\mE_\mr(\cdot)$ at $R$, i.e., $\partial \mE_\mr(R)$, satisfies\footnote{As usual, for a given set $S$, $\textnormal{conv}(S)$ denotes the \emph{convex hull} of $S$.} 
\begin{equation}
\partial\mE_\mr(R) = \textnormal{conv}\left( \left\{ \left.\frac{\partial \mE_\mr(a, Q)}{\partial a }\right|_{a = R} : \mE_{\mr}(R,Q)= \mE_\mr(R) \right\}\right). 
\label{eq:thrm3-i}
\end{equation}
 \item[(ii)] Define $\rho^\ast_R \eqdef \max \left\{ |\rho^\ast| : \rho^\ast \in \partial \mE_\mr(R) \right\}$. If there exists $Q \in \cP(\cX)$ such that $\mE_\mr(R,Q) = \mE_\mr(R)$, $(Q,W) \in \cP_{\textnormal{ns}}$ and $ -\left.\frac{\partial \mE_\mr(a, Q)}{\partial a }\right|_{a = R} = \rho^\ast_R$, then there exists an $(N,R)$ code $(f, \varphi)$ and $K_5 \in \bbR^+$ that depends on $W,R$ and $Q$ such that 
\begin{equation}
\mP_{\me}(f, \varphi) \leq \frac{K_5}{N^{0.5(1+\rho_R^\ast)}}e^{-N \mE_\mr(R)}, 
\label{eq:thrm3-ii}
\end{equation}
for all $N \in \bbZ^+$. 

\item[(iii)] If $W$ satisfies the condition
\begin{equation}
W(y|x)>0, \textnormal{ for all }(x,y) \in \cX \times \cY, \label{eq:thrm3-iii}
\end{equation}
then for any $Q \in \cP(\cX)$ with $\mE_{\mr}(R,Q) = \mE_{\mr}(R)$, $(Q,W) \in \cP_{\textnormal{ns}}$. Hence, \eqref{eq:thrm3-iii} is a sufficient condition for the existence of a $Q \in \cP(\cX)$ as in item (ii) above. \eot
\end{itemize}
\label{thrm:thrm3}
\end{theorem}
\begin{IEEEproof}
Theorem~\ref{thrm:thrm3} is proved in Section~\ref{app-thrm3}.
\end{IEEEproof}

\begin{remark}
\begin{itemize}
\item[(i)] It is evident that $\rho^\ast_R$, as defined in item (ii) of Theorem~\ref{thrm:thrm3}, is the absolute value of the left derivative of $\mE_\mr(\cdot)$ at $R$. Further, it is worth noting that in \cite{altug12-e}, the authors proved that for any $W \in \cP(\cY|\cX)$ with\footnote{See \cite[pg.~158]{gallager68} for the definition of $R_\infty$.} $R_\infty < C$, and $R_\infty < R < C$ and $\epsilon >0$, the maximum error probability of any constant composition $(N,R)$ code is lower bounded by 
\begin{equation}
\label{eq:sp}
\frac{K_5 e^{-N\mE_{\textrm{SP}}(R)}}{N^{\frac{1}{2}\left( 1 + \epsilon + \tilde{\rho}^{\ast}_R \right)}}, 
\end{equation}
for all sufficiently large $N$, where $K_5$ is a positive constant that depends on $W$, $R$ and $\epsilon$, $\mE_{\mSP}(R)$ is the \emph{sphere-packing exponent} (e.g., \cite[Eq.~(5.6.2)]{gallager68}), and $ \tilde{\rho}^{\ast}_R$ is the absolute value of the left derivative of $\mE_\textrm{SP}(\cdot)$ at $R$. For $R_\mcr < R < C $, $\mE_{\mSP}(R) = \mE_{\mr}(R)$ (e.g., \cite[pg.~160]{gallager68}) and also\footnote{Since the non-increasing and convex curves $\mE_{\mSP}(\cdot)$ and $\mE_{\mr}(\cdot)$ agree on an interval around $R$, the maximum magnitude of their subdifferentials at $R$ are also equal.} $\tilde{\rho}^{\ast}_R = \rho_R^\ast $. 

\item[(ii)] In \cite{Dobrushin62}, Dobrushin considers a strongly symmetric channel\footnote{A channel is strongly symmetric if every row (resp. column) is a permutation of every other row (resp. column).} with $R_{\mcr} < C$ and proves\footnote{The English translation of this work mistakenly states the pre-factor as $O(N^{-\frac{1}{2(1 + |\mE_{\mr}^\prime(R)|)}})$. We thank Jonathan Scarlett for pointing out the fact that the original Russian version has the following correct form.} the existence  of an $(N,R)$ code $(f, \varphi)$ such that $\mP_{\me}(f, \varphi) \leq O(N^{-0.5(1+|\mE_\mr^\prime(R)|)})e^{-N \mE_{\mr}(R)}$. One can verify\footnote{For contradiction, assume $(U_\cX,W) \in \cP_{\textnormal{s}}$, which, due to the strong symmetry of $W$, implies that there exists a positive constant $c$ such that $W(y|x) \in \{ 0, c\}$ for all $(x,y) \in \mathcal{X} \times \mathcal{Y}$. The last observation implies that the mutual information random variable, i.e., $\log \frac{W(y|x)}{\sum_z U_\cX(z)W(y|z)}$, has zero variance, which, in turn, implies that (e.g., \cite[pg.~160]{gallager68}) $R_\mcr = C$, by noticing the fact that (e.g., \cite[Thrm.~4.5.2]{gallager68}) $U_\cX$ is a capacity achieving input distribution for $W$.} that for any strongly symmetric channel, say $W$, with $R_{\mcr} < C$, $(U_\cX,W) \in \cP_{\textnormal{ns}}$, where $U_\cX$ is the uniform distribution over the input alphabet $\cX$. Since $U_\cX$ attains $\mE_{\mr}(R)$ for all $R \in [R_{\mcr},C]$ (e.g., \cite[pg.~145]{gallager68}), $\mE_{\mr}^\prime(\cdot,U_\cX) = \mE_{\mr}^\prime(\cdot)$ over $(R_{\mcr},C)$ and hence we conclude that item (ii) of Theorem~\ref{thrm:thrm3} is a generalization of the aforementioned  result in \cite{Dobrushin62}.\eor
\end{itemize}
\label{rem:rem1.5}
\end{remark}

Singularity is also crucial regarding the pre-factor of the ensemble average error probability for rates below the critical rate.

\begin{theorem}
Let $W \in \cP(\cY|\cX)$ be arbitrary with $C >0$ and $R \leq R_\mcr$. 
\begin{itemize}
\item[(i)] If the pair $(Q,W)$ is singular and $\mE_{\mo}(1,Q) = \max_{P \in \cP(\cX)} \mE_{\mo}(1,P)$, then
\begin{equation}
K_6 e^{-N\mE_{\mr}(R)}\leq \bar{\mP}_\me(Q,N,R) \leq  e^{-N\mE_{\mr}(R)},
\label{eq:thrm2-1}
\end{equation}
for any $N \in \bbZ^+$ and for some $0<K_6 < 1$ that depends on $W,R$ and $Q$.

\item[(ii)](Gallager~\cite{gallager73}) If the pair $(Q,W)$ is nonsingular and $\mE_{\mo}(1,Q) = \max_{P \in \cP(\cX)} \mE_{\mo}(1,P)$, then  
\begin{equation}
\bar{\mP}_{\me}(Q, N, R) \sim \frac{g}{\sqrt{N}}e^{-N\mE_{\mr}(R)},
\label{eq:thrm2-2}
\end{equation}
where $g$ is a positive constant that is explicitly characterized in \cite{gallager73}. \eot
\end{itemize}
\label{thrm:thrm2}
\end{theorem}
\begin{IEEEproof}
Theorem~\ref{thrm:thrm2} is proved in Section~\ref{sec:proof-thrm2}. 
\end{IEEEproof}

\begin{remark}
\begin{itemize}
\item[(i)] Theorem~\ref{thrm:thrm2} corrects a small oversight in \cite{gallager73}, which asserts the conclusion in (ii) for all channels.
In fact, the statement and the proof given there only hold in the nonsingular case \cite{gallager12}. 
\item[(ii)] The abrupt drop in the order of the pre-factor at $R_\mcr$ highlights a previously unreported role that the critical rate plays in the random coding bound. \eor
\end{itemize}
\label{rem:rem2}
\end{remark}

\section{Proof of Theorem~\ref{thrm:thrm1}}
\label{sec:proof-thrm1}
\subsection{Overview}
\label{ssec:proof-thrm1-overview}
From the well-known random coding arguments (e.g., \cite[pg.~136]{gallager68}) one can deduce that for any message $m$
\begin{equation}
\bar{\mP}_{\me}(Q,N,R) \leq \sum_{\bx_m, \by}Q(\bx_m)W(\by|\bx_m) \Pr \left\{ \bigcup_{m^\prime \neq m} \left\{ \log \frac{W(\by|\bx_m)}{W(\by|\bX_{m^\prime})} \leq 0 \right\} \right\}. 
\label{eq:overview1}
\end{equation}
For the sake of notational convenience, let $\mathcal{E}_m \eqdef \bigcup_{m^\prime \neq m} \left\{ \log \frac{W(\bY|\bX_m)}{W(\bY|\bX_{m^\prime})} \leq 0 \right\}$ denote the error event conditioned on message $m$. 

One obvious way to relax the right side of \eqref{eq:overview1} to make it more tractable is to use the union bound. A straightforward application of the union bound is loose, however, because some realizations of $\bX_m$ and $\bY$ are such that $\left\{ \log \frac{W(\bY|\bX_m)}{W(\bY|\bX_{m^\prime})} \leq 0 \right\}$ is likely to occur for many $m^\prime$. One standard workaround is to define a set of ``bad'' $\bX_m$ and $\bY$ realizations $\cD_N \in \cX^N \times \cY^N$ and proceed as follows
\begin{align}
\bar{\mP}_{\me}(Q,N,R) & \leq \Pr(\mathcal{E}_m \cap \cD_N) + \Pr(\mathcal{E}_m \cap \cD_N^c) \nonumber \\
 & \leq \Pr(\cD_N) +  (\lceil e^{NR} \rceil -1 )\Pr\left\{ \cD_N^c \cap \left\{ \log \frac{W(\bY|\bX)}{W(\bY|\bZ)} \leq 0\right\} \right\}.  
\label{eq:overview2}
\end{align}

\begin{remark}
\begin{itemize}
\item[(i)] Equation \eqref{eq:overview2} is due to Fano~\cite[pg.~307, Theorem]{fano61} and is valid for any auxiliary set $\cD_N$, where $X,Y$ and $Z$ are distributed with $P_{X,Y,Z}(x,y,z) = Q(x)W(y|x)Q(z)$. Fano provides a choice of $\cD_N$ for which a large deviations analysis of the right side of \eqref{eq:overview2} yields the random coding exponent. 

\item[(ii)] 
\label{item:LDPC} It is evident that the introduction of an auxiliary set in Fano's bound is not limited to random code ensembles, but can also be employed to analyze the error probability of a given block code under maximum likelihood decoding. In particular, Gallager used this idea in his analysis of low-density parity-check (LDPC) codes~\cite[Section~3.3]{gallager63}. After the invention of turbo codes \cite{turbo93} and the rediscovery of LDPC codes \cite{LDPC97}, there has been considerable interest in deriving  efficiently computable bounds on the performance of a given block code (e.g., \cite{shulman-feder99}--\cite{shamai2009} and references therein). Researching these bounds for possible refinements, in particular characterizing the pre-factors of the exponentially vanishing terms, is an interesting future research direction, which is not pursued in this paper. 

\item[(iii)] There are other ways to control the aforementioned loss. One alternative is to use the following bound by Gallager (e.g., \cite[eq.~(5.6.7)]{gallager68})
\begin{equation}
\bar{\mP}_{\me}(Q,N,R) \leq \sum_{\bx_{m}, \by}Q(\bx_{m})W(\by|\bx_{m})\left(\sum_{m^\prime \neq m} \Pr\left\{ \log \frac{W(\by|\bx_m)}{W(\by|\bX_{m^\prime})} \leq 0\right\}    \right)^\rho,
\label{eq:overview4}
\end{equation}
for any $\rho \in [0,1]$. Although the bound in \eqref{eq:overview4} is sufficient to obtain the random coding exponent, the bound in \eqref{eq:overview2} seems to be better suited to obtaining improved pre-factors. 

A tighter alternative  to \eqref{eq:overview2} is (e.g., \cite[pg.~137]{gallager68}, \cite[Theorem~16]{polyanskiy10})
\begin{equation}
\bar{\mP}_{\me}(Q,N,R) \leq \sum_{\bx, \by}Q(\bx)W(\by|\bx) \min\left\{ 1, (\lceil e^{NR}\rceil - 1)\Pr\left\{ \log \frac{W(\by|\bx)}{W(\by|\bZ)} \leq 0\right\} \right\}.
\label{eq:overview3}
\end{equation}
The alternative proof of Theorem~\ref{thrm:thrm1} by Scarlett \emph{et al.}~\cite{scarlett13}, mentioned earlier, uses this bound as its starting point.
Although their derivation is simpler than the one given here based on \eqref{eq:overview2}, the latter has the merit of being the starting point for possible refinements of error probability bounds for a given block code, as noted above.       
\eor
\end{itemize}
\label{rem:overview1}
\end{remark}

Next, one needs to choose an appropriate $\cD_N$ and upper bound the terms on the right side of \eqref{eq:overview2}. Our choice will essentially be Fano's choice for $\cD_N$ and our analysis will vary depending on whether the pair $(Q,W)$ is singular. Specifically, if the pair $(Q,W)$ is singular, then we use Fano's choice. However, if the pair $(Q,W)$ is nonsingular, then a perturbed version of Fano's $\cD_N$ gives a better pre-factor and we will use this perturbed version. 

Before proceeding further, we note the following useful facts that will be used throughout the paper.
\begin{lemma}
Let $W \in \cP(\cY|\cX)$ be arbitrary with $R_{\mcr} < C$. Fix any $Q \in \cP(\cX)$ such that $\mE_{\mr}(R,Q) >0$ for some $R > R_\infty$.
\begin{itemize}
\item[(i)] $\frac{\d^2 \mE_{\mo}(\rho,Q)}{\d \rho^2} < 0$ for all $\rho \in \bbR_+$. 

\item[(ii)] For any $R_\infty < r \leq \mI(Q;W) $, there exists a unique $\rho_r^\ast(Q) \in \bbR_+$ such that 
\begin{equation}
\sup_{\rho \in \bbR_+} \left\{ - \rho r + \mE_{\mo}(\rho, Q) \right\} = -\rho_r^\ast(Q) r + \mE_{\mo}(\rho_r^\ast(Q), Q). \label{eq:lem-prelim-1}
\end{equation}
Further, $\rho_r^\ast(Q) \in \bbR_+$ is the unique number satisfying 
\begin{equation}
\left.\frac{\partial \mE_{\mo}(\rho, Q)}{\partial \rho}\right|_{\rho = \rho_r^\ast(Q)}=r.
\label{eq:lem-prelim-2}
\end{equation}

\item[(iii)] $\rho_r^\ast(Q) \in (0,1)$ if and only if $r \in \left( \left. \frac{\partial \mE_{\mo}(\rho, Q)}{\partial \rho}\right|_{\rho =1}, \mI(Q;W)\right)$.

\item[(iv)] $\rho^\ast_{(\cdot)}(Q)$ is continuous over $\left( \left. \frac{\partial \mE_{\mo}(\rho, Q)}{\partial \rho}\right|_{\rho =1}, \mI(Q;W)\right)$ and on this interval, satisfies 
\begin{equation}
\rho_r^\ast(Q) = - \left.\frac{\partial \mE_{\mr}(a,Q)}{\partial a}\right|_{a =r}. 
\label{eq:lem-prelim-3}
\end{equation}
\end{itemize}
\eot
\label{lem:prelim}
\end{lemma}


\begin{IEEEproof}
The proof is given in Appendix~\ref{app:lem-prelim-pf}. 
\end{IEEEproof}

To define the auxiliary set, we need the following definitions. First, fix some $W \in \cP(\cY|\cX)$ with $R_{\mcr} < C$. Consider some $Q \in \cP(\cX)$ and $R \in \bbR_+$ such that $R_\mcr(Q) < R < \mI(Q;W)$. Define
\begin{equation}
P_{X,Y,Z}(x,y,z) \eqdef Q(x)W(y|x)Q(z),
\label{eq:P_XYZ}
\end{equation}
for all $(x,y,z) \in \cX \times \cY \times \cX$. Also, let
\begin{equation}
\tilde{P}_{X,Y,Z}(x,y,z) \eqdef 
\begin{cases}
\frac{P_{X,Y,Z}(x,y,z)}{P_{X,Y,Z}\left\{ \tilde{\cS}_{Q}\right\}} & \textrm{ if } (x,y,z) \in \tilde{\cS}_{Q}, \\
0 & \textrm{ else}.
\end{cases}
\label{eq:tP_XYZ}
\end{equation}
Let $P^{N}_{X,Y,Z}(\bx, \by, \bz) \eqdef \prod_{n=1}^{N}P_{X,Y,Z}(x_{n}, y_{n}, z_{n})$ and $\cS_{Q}^{N}$ (resp. $\tilde{S}_{Q}^{N}$) denote the $N$-fold cartesian product of $\cS_{Q}$ (resp. $\tilde{\cS}_{Q}$). Hence, 
\[
P_{X,Y,Z}^{N}\left\{ \bx, \by, \bz | \tilde{\cS}_{Q}^{N} \right\} = \tilde{P}^{N}_{X,Y,Z}(\bx, \by, \bz) \eqdef \prod_{n=1}^{N}\tilde{P}_{X,Y,Z}(x_{n}, y_{n}, z_{n}).
\]

For any $\rho \in [0,1]$, let\footnote{The following two quantities are defined for any $\rho \in \bbR_+$ in items (i) and (v) of Definition~\ref{def:def1} in Appendix~\ref{app-prelim}, respectively. We reproduce them here for the reader's convenience.} 
\begin{align}
f_\rho(y) & \eqdef \frac{\left[ \sum_{x \in \cX}Q(x)W(y|x)^{1/(1+\rho)}\right]^{1+\rho}}{\sum_{b \in \cY}\left[ \sum_{a \in \cX}Q(a)W(b|a)^{1/(1+\rho)}\right]^{1+\rho}}, \forall \, y \in \cY. \label{eq:f-star} \\
\Lambda_\rho\left( \lambda \right) & \eqdef \log \mE_{P_{X,Y}}\left[ e^{\lambda \log \frac{f_\rho(Y)}{W(Y|X)}}\right], \forall \, \lambda \in \bbR. 
\label{eq:lambda-N} 
\end{align}

For any $\rho \in [0,1]$, $\log\frac{f_\rho(y)}{W(y|x)} \in \bbR$ for all $(x,y) \in \cS_Q$, hence $\Lambda_\rho(\cdot)$ is infinitely differentiable on $\bbR$. Thus, for any $\rho \in [0,1]$, the following is well-defined
\begin{equation}
D_{\mo}(\rho) \eqdef \Lambda^\prime_\rho\left( \frac{\rho}{1+\rho}\right). 
\label{eq:Do-N}
\end{equation}

Let $\{ \epsilon_N\}_{N \geq 1}$ be a sequence of nonnegative real numbers such that $\lim_{N \rightarrow \infty} \epsilon_N = 0$ and define $R_N \eqdef R - \epsilon_N$. Let $N \in \bbZ^+$ be sufficiently large such that $R_N > R_\mcr(Q)$. For the sake of notational convenience, let
\begin{equation}
\rho^\ast_N \eqdef \rho_{R_N}^\ast(Q) = -\left.\frac{\d \mE_\mr(r,Q)}{\d r}\right|_{r = R_N},
\label{eq:rho-star-Q}
\end{equation} 
whose existence is ensured by \eqref{eq:lem-prelim-3}. 

We finally define the auxiliary set as follows:
\begin{equation}
\mathcal{D}_{N}(\epsilon_N) \eqdef \left\{ \frac{1}{N} \sum_{n=1}^{N}\log\frac{f_{\rho^\ast_N}(Y_{n})}{W(Y_{n}|X_{n})} > D_{\mo}(\rho^\ast_N)\right\}. \label{eq:cDo-N}
\end{equation}

Using the particular set defined in \eqref{eq:cDo-N}, equation \eqref{eq:overview2} reads  
\begin{align}
\bar{\mP}_{\me}(Q,N,R) & \leq P_{X,Y}^{N}\left\{ \cD_{N}(\epsilon_N) \right\} \nonumber \\
& \quad + (\lceil e^{NR} \rceil - 1) P_{X,Y,Z}^{N}\left\{ \frac{1}{N}\sum_{n=1}^{N}\log\frac{f_{\rho^\ast_N}(Y_{n})}{W(Y_{n}|X_{n})}\leq D_{\mo}(\rho^\ast_N), \frac{1}{N}\sum_{n=1}^{N}\log\frac{W(Y_{n}|X_{n})}{W(Y_{n}|Z_{n})} \leq 0 \right\}. \label{eq:thrm1-pf-step1-1.0}
\end{align}

\begin{remark}
\begin{itemize}
\item[(i)] Setting $\epsilon_N =0$ for all $N \in \bbZ^+$ gives Fano's choice of the auxiliary set. After this point, he proceeds with Chernoff bound arguments to upper bound the right side of \eqref{eq:thrm1-pf-step1-1.0} to deduce the random coding upper bound\footnote{Fano's exponent, $\mE_{\textrm{F}}(\cdot)$ (e.g., item (iv) of Definition~\ref{def:def1} in Appendix~\ref{app-prelim}) has a different form than $\mE_\mr(\cdot)$, yet they can be shown to be equal (e.g., Lemma~\ref{lem:lem3} in Appendix~\ref{app-prelim}).} with a pre-factor of $O(1)$ \cite[pp.~324--331]{fano61}.

\item[(ii)] If $(Q,W)$ is nonsingular, then one can simply replace 
Fano's use of the Chernoff bound with some scalar and vector exact asymptotic results (e.g., \cite{bahadur-rao60}, \cite{ney83}) to obtain a bound with the same exponent and a 
pre-factor of $O(1/\sqrt{N})$ \cite{altug12-b}.
Moreover, $O(1/\sqrt{N})$ is the tightest pre-factor possible if $\epsilon_N =0$ in the sense that one can show that $P_{X,Y}^N\left\{ \cD_N(\epsilon_N) \right\} \sim \Theta(1/\sqrt{N}) e^{-N\mE_{\mr}(R,Q)}$.

\item[(iii)] If $(Q,W)$ is nonsingular, then setting $\epsilon_N = 0$ for all $N \in \bbZ^+$ is not the best choice. Indeed, with this choice, one can prove an upper bound of $O(1/N)e^{-N\mE_{\mr}(R,Q)}$ on the second term of \eqref{eq:thrm1-pf-step1-1.0}, using the fact that
\begin{equation}
\left[\log\frac{f_{\rho^\ast_N}(Y)}{W(Y|X)},   \log\frac{W(Y|X)}{W(Y|Z)}\right]^T,
\label{eq:rem-thrm1-pf-1}
\end{equation}
is nonsingular when it is distributed according to $\tilde{P}_{X,Y,Z}$, i.e., the covariance matrix of this random vector under $\tilde{P}_{X,Y,Z}$ is nonsingular. This follows from the nonsingularity of $(Q,W)$. For the first term,
one obtains a bound of $O(1/\sqrt{N}) e^{-N\mE_{\mr}(R,Q)}$. Thus, the
pre-factor is dominated by the first term, and it is advantageous to 
increase $\epsilon_N$ to decrease the first term at the expense of
the second. In Section~\ref{ssec:proof-thrm1-i} we shall see how to
choose $\epsilon_N$ in order to equalize the order of the two 
pre-factors.

\item[(iv)] If $(Q,W)$ is singular, then $\log \frac{W(Y|X)}{W(Y|Z)} = 0 $, $\tilde{P}_{X,Y,Z}-\textrm{(a.s.)}$. Hence, the random vector given in \eqref{eq:rem-thrm1-pf-1} is singular when it is distributed with $\tilde{P}_{X,Y,Z}$, i.e., the covariance matrix of this random vector under $\tilde{P}_{X,Y,Z}$ is singular. Therefore, we expect to have an upper bound on the second term of \eqref{eq:thrm1-pf-step1-1.0} with an $O(1/\sqrt{N})$ pre-factor and hence we will set $\epsilon_N =0$ for all $N \in \bbZ^+$ for this case. The details of the derivation is given in Section~\ref{ssec:proof-thrm1-ii}.  

\item[(v)] As seen in items (iii) and (iv) above, whether $(Q,W)$ satisfies \eqref{eq:singular-defn} is closely related to the singularity of the covariance matrix of the random vector in \eqref{eq:rem-thrm1-pf-1} under $\tilde{P}_{X,Y,Z}$. This relation is our rationale for calling Definition~\ref{def:singular} \emph{singularity}. \eor
\end{itemize}
\label{rem:rem-thrm1-pf-1}
\end{remark}

Before proceeding further, we define the following quantities 

For any $\rho \in [0,1]$, $\lambda \in \bbR$ and $\bv \in \bbR^2$, 
\begin{equation}
\tilde{P}_{X,Y}^{\lambda,\rho}(x,y) \eqdef 
\begin{cases}
\frac{Q(x)W(y|x)^{1-\lambda}f_{\rho}(y)^\lambda}{\sum_{(a,b) \in \cS_Q}Q(a)W(b|a)^{1-\lambda}f_\rho(b)^\lambda}, &  \textrm{ if } (x,y) \in \cS_Q, \\
0, & \textrm{ else}.
\end{cases}
\label{eq:lem-pos-variance-pf1}
\end{equation}
\begin{equation}
\Lambda_{1, \rho}(\bv) \eqdef \log \mE_{\tilde{P}_{X,Y,Z}}\left[ e^{\bv_{1}\log \frac{W(Y|X)}{f_\rho(Y)} + \bv_{2}\log \frac{W(Y|Z)}{W(Y|X)}}\right]. 
\label{eq:lambda1}
\end{equation}
Clearly, $\tilde{P}_{X,Y}^{\lambda, \rho}$ is a well-defined probability measure and $\Lambda_{1, \rho}(\cdot)$ is infinitely differentiable on $\bbR^2$. Further, 
\begin{lemma}
Fix an arbitrary $r \in \left(R_\mcr(Q), \mI(Q;W) \right)$. Let $\rho \eqdef -\left. \frac{\partial \mE_\mr(a,Q)}{\partial a}\right|_{a = r} \in (0,1)$ and $\tilde{\bv} \eqdef \left[ \frac{1-\rho}{1+\rho}, \frac{1}{1+\rho}\right]^T$. We have 
\begin{itemize}
\item[(i)]
\begin{equation}
\left[ \left.\frac{\d \Lambda_{1, \rho}(\bv_1, \tilde{\bv}_2)}{\d \bv_1}\right|_{\bv_1 = \tilde{\bv}_1}, \left.\frac{\d \Lambda_{1, \rho}(\tilde{\bv}_1, \bv_2)}{\d \bv_2}\right|_{\bv_2 = \tilde{\bv}_2}\right]^T  =  [-\Lambda_\rho^\prime(\rho/(1+\rho)),0]^T. \label{eq:tilted-mean2-1} 
\end{equation}
\item[(ii)]
\begin{equation}
\Lambda_{1, \rho}(\tilde{\bv}) = -\log P_{X,Y,Z}\left\{ \tilde{\cS}_Q \right\} + 2 \Lambda_{\rho}\left( \frac{\rho}{1+\rho}\right).
\label{eq:thrm1-pf-step5-8.1}
\end{equation}
\end{itemize}
\eot
\label{lem:diff-new}
\end{lemma}

\begin{IEEEproof}
The proof is given in Appendix~\ref{app-lem-diff-new-pf}.
\end{IEEEproof}

\subsection{Proof of item (i) of Theorem~\ref{thrm:thrm1}}
\label{ssec:proof-thrm1-ii} 

Assume the pair $(Q,W)$ is singular. As pointed out in item (iii) of Remark~\ref{rem:rem-thrm1-pf-1}, we use the quantities given in Section~\ref{ssec:proof-thrm1-overview} with $\epsilon_N =0$ for all $N \in \bbZ^+$. Specifically, define 
\begin{equation}
\rho^\ast \eqdef  - \left.\frac{\partial \mE_\mr(r, Q)}{\partial r}\right|_{r= R}.
\label{eq:singular-rho}
\end{equation}
Let $f^\ast$, $\Lambda(\cdot)$ and $D_\mo$ denote the quantities defined in \eqref{eq:f-star}, \eqref{eq:lambda-N} and \eqref{eq:Do-N}, respectively, by choosing $\rho = \rho^\ast$. For convenience, let $\cD_N$ denote the set defined in \eqref{eq:cDo-N} with the aforementioned choices. Particularizing \eqref{eq:overview2}, we have 
\begin{equation}
\bar{\mP}_{\me,m}(Q, N,R) \leq P_{X,Y}^{N}\left\{ \cD_{N} \right\} + (\lceil e^{NR}  \rceil -1) P_{X,Y,Z}^{N}\left\{ \frac{1}{N}\sum_{n=1}^{N}\log \frac{f^{\ast}(Y_{n})}{W(Y_{n}|X_{n})} \leq D_{\mo}, \frac{1}{N}\sum_{n=1}^{N} \log \frac{W(Y_{n}|X_{n})}{W(Y_{n}|Z_{n})} \leq 0 \right\}. 
\label{eq:irreg-bound-1}
\end{equation}

We begin by deriving an upper bound on the first term in the right side of \eqref{eq:irreg-bound-1}. 

\begin{lemma}
$\Lambda^{\prime \prime}(\lambda) >0$, for all $\lambda \in \bbR$. \eot
\label{lem:irreg-lem1}
\end{lemma}

\begin{IEEEproof}
The proof proceeds by contradiction. One can check that
\begin{equation}
\left[ \exists \, \lambda \in \bbR \textrm{ with } \Lambda^{\prime \prime}(\lambda) =0 \right] \Longleftrightarrow \left[ \log \frac{f^\ast(Y)}{W(Y|X)} = \Lambda^\prime(\lambda), \, P_{X,Y}-\textrm{(a.s.)}\right]. \label{eq:lem-irreg-lem-1-pf1}
\end{equation}
Further, define $\tilde{\cY} \eqdef \{y \in \cY : \cX_y \neq \emptyset \}$. Note that $\tilde{\cY} \neq \emptyset$. Since the pair $(Q,W)$ is singular, for some $\delta_y \in \bbR^+$
\begin{equation}
W(y|x) = \delta_y, \, \forall x \in \cX_y, 
\label{eq:lem-irreg-lem-1-pf1.5}
\end{equation}
which, in turn, implies that
\begin{equation}
f^\ast(y) = \frac{\delta_y Q\left\{\cX_y\right\}^{1+\rho^\ast}}{\sum_{b \in \tilde{\cY}}\delta_b Q\left\{\cX_b\right\}^{1+\rho^\ast}}. \label{eq:lem-irreg-lem-1-pf2}
\end{equation}
Equations \eqref{eq:lem-irreg-lem-1-pf1.5} and \eqref{eq:lem-irreg-lem-1-pf2} imply that 
\begin{equation} 
\log \frac{f^\ast(y)}{W(y|x)} = \log \frac{Q\left\{\cX_y\right\}^{1+\rho^\ast}}{\sum_{b \in \tilde{\cY}}\delta_b Q\left\{\cX_b\right\}^{1+\rho^\ast}}, \, \forall \, (x,y) \in \tilde{\cS}_Q. \label{eq:lem-irreg-lem-1-pf3}
\end{equation}

Due to \eqref{eq:lem-irreg-lem-1-pf3}, one can check that the right side of \eqref{eq:lem-irreg-lem-1-pf1} is equivalent to saying that $Q\left\{ \cX_y \right\}$ is constant for all $y \in \tilde{\cY}$. This last observation, coupled with the singularity of the pair $(Q,W)$, further implies that
\begin{equation}
\mE_\mo(\rho, Q) = -(1+\rho)\log Q\left\{ \cX_y \right\} - \log \sum_y \delta_y, \label{eq:lem-irreg-lem-1-pf4}
\end{equation}
for all $\rho \in \bbR_+$. Evidently, \eqref{eq:lem-irreg-lem-1-pf4} implies that $\frac{\d^2 \mE_\mo(\rho, Q)}{\d \rho^2} =0$, for all $\rho \in \bbR_+$, which contradicts item (i) of Lemma~\ref{lem:prelim}.
\end{IEEEproof}

Equipped with Lemma~\ref{lem:irreg-lem1}, we can apply Lemma~\ref{lem:EA-new} in Appendix~\ref{app-rv-EA} to obtain\footnote{In the conference version of this work, the second term in the braces of \eqref{eq:singular-error} (resp. \eqref{eq:irreg-bound-part2-3}) is incorrectly written as $\frac{1}{\sqrt{2 \pi}\eta}$ \cite[Eq.~(66)]{altug-allerton12} (resp. $\frac{1}{\sqrt{2 \pi} \tilde{\eta}}$ \cite[Eq.~(69)]{altug-allerton12}). The correct form is $\frac{1}{\sqrt{2 \pi \Lambda^{\prime \prime}(\eta)}\eta}$ (resp. $\frac{1}{\sqrt{2 \pi \Lambda_{\mo}^{\prime \prime}(\tilde{\eta})}\tilde{\eta}}$), as given in \eqref{eq:singular-error} (resp. \eqref{eq:irreg-bound-part2-3}).}
\begin{equation}
P_{X,Y}^N\left\{ \cD_N \right\} \leq e^{-N \Lambda^\ast(D_\mo)}\frac{1}{\sqrt{N}}\left\{ \frac{m_3}{\Lambda^{\prime \prime}(\eta)^{3/2}} + \frac{1}{\sqrt{2 \pi \Lambda^{\prime \prime}(\eta)}\eta} \right\}, \label{eq:singular-error}
\end{equation}
where $\eta \eqdef \frac{\rho^\ast}{1+\rho^\ast}$, $m_3 \eqdef \mE_{\tilde{P}_{X,Y}^{\eta, \rho^\ast}}\left[ \left| \log \frac{f^\ast(Y)}{W(Y|X)} - \Lambda^\prime(\eta) \right|^3  \right]$ with $\tilde{P}_{X,Y}^{\eta, \rho^\ast}$ as defined in \eqref{eq:lem-pos-variance-pf1}, and $\Lambda^\ast(D_\mo)$ is the Fenchel-Legendre transform of $\Lambda(\cdot)$ at $D_\mo$, i.e.,
\begin{equation}
\Lambda^\ast(D_\mo) \eqdef \sup_{\lambda \in \bbR}\left\{ D_\mo \lambda - \Lambda(\lambda) \right\}.
\label{eq:fenchel-1}
\end{equation} 

Since $\Lambda(\cdot)$ is convex, the definition of $D_\mo$ and \eqref{eq:fenchel-1} imply that 
\begin{equation}
\Lambda^\ast(D_\mo) = \eta \Lambda^\prime(\eta) - \Lambda(\eta). 
\label{eq:irreg-bound-part1-exp1}
\end{equation}

Moreover, Lemma~\ref{lem:lem3} and \eqref{eq:lem5-2} in Appendix~\ref{app-prelim} imply that 
\begin{equation}
\mE_\mr(R,Q) = \eta \Lambda^\prime(\eta) - \Lambda(\eta).
\label{eq:irreg-bound-part1-exp1.5}
\end{equation}

By substituting \eqref{eq:irreg-bound-part1-exp1.5} into \eqref{eq:irreg-bound-part1-exp1}, we deduce that  
\begin{equation}
\Lambda^\ast(D_\mo) = \mE_\mr(R,Q), 
\label{eq:irreg-bound-part1-exp1.75}
\end{equation}
which, in turn, implies that 
\begin{equation}
P_{X,Y}^N\left\{ \cD_N \right\} \leq e^{-N \mE_\mr(R,Q)}\frac{1}{\sqrt{N}}\left\{ \frac{m_3}{\Lambda^{\prime \prime}(\eta)^{3/2}} + \frac{1}{\sqrt{2 \pi \Lambda^{\prime \prime}(\eta)}\eta} \right\}.
\label{eq:irreg-bound-part1}
\end{equation}

In order to upper bound the remaining term in the right side of \eqref{eq:irreg-bound-1}, we first note that 
\begin{align}
\beta_N & \eqdef P_{X,Y,Z}^{N}\left\{ \frac{1}{N}\sum_{n=1}^{N}\log\frac{f^{\ast}(Y_{n})}{W(Y_{n}|X_{n})} \leq D_{\mo}, \frac{1}{N}\sum_{n=1}^{N}\log\frac{W(Y_{n}|X_{n})}{W(Y_{n}|Z_{n})} \leq 0 \right\} \nonumber \\
 & = P^N_{X,Y,Z}\left\{ \tilde{S}^N_Q\right\} \tilde{P}_{X,Y,Z}^{N}\left\{ \frac{1}{N}\sum_{n=1}^{N}\log\frac{W(Y_{n}|X_{n})}{f^{\ast}(Y_{n})} \geq -D_{\mo}, \frac{1}{N}\sum_{n=1}^{N}\log\frac{W(Y_{n}|Z_{n})}{W(Y_{n}|X_{n})} \geq 0 \right\} \nonumber \\
 & = P^N_{X,Y,Z}\left\{ \tilde{S}^N_Q\right\} \tilde{P}_{X,Y,Z}^{N}\left\{ \frac{1}{N}\sum_{n=1}^{N}\log\frac{W(Y_{n}|X_{n})}{f^{\ast}(Y_{n})} \geq -D_{\mo} \right\}, \label{eq:irreg-bound-part2-1}
\end{align}
where \eqref{eq:irreg-bound-part2-1} follows by noting $\log \frac{W(y|z)}{W(y|x)} = 0$ for all $(x,y,z) \in \tilde{S}_Q$, which is a direct consequence of the singularity of the pair $(Q,W)$. 

Next, define
\begin{equation}
\forall \, \lambda \in \bbR, \, \Lambda_\mo(\lambda) \eqdef \log \mE_{\tilde{P}_{X,Y,Z}}\left[ e^{\lambda \log \frac{W(Y|X)}{f^\ast(Y)}}\right], 
\label{eq:irreg-Lambda-0}
\end{equation}
and note that $\Lambda_\mo(\cdot)$ is infinitely differentiable on $\bbR$. Moreover, one can  
check that 
\begin{equation}
\forall \, \bv \in \bbR^2, \, \Lambda_{\mo}(\bv_1) = \Lambda_1(\bv), \label{eq:irreg-Lambda0-Lambda1}
\end{equation}
where $\Lambda_1(\cdot)$ denotes $\Lambda_{1, \rho^\ast}(\cdot)$ (e.g., \eqref{eq:lambda1}) for notational convenience. Further, for any $\lambda \in \bbR$, define 
\begin{equation}
\tilde{Q}_{X,Y,Z}^\lambda(x,y,z) \eqdef 
\begin{cases}
\frac{\tilde{P}_{X,Y,Z}(x,y,z)W(y|x)^\lambda f^\ast(y)^{-\lambda}}{\sum_{(a,b,c) \in \tilde{\cS}_Q}\tilde{P}_{X,Y,Z}(a,b,c)W(b|a)^\lambda f^\ast(b)^{-\lambda}} & \textrm{ if } (x,y,z) \in \tilde{\cS}_Q, \\
0 & \textrm{ else.}
\end{cases}
\label{eq:irreg-tilted-Q}
\end{equation}
It is evident that $\tilde{Q}_{X,Y,Z}^\lambda$ is a well-defined probability measure and equivalent to $\tilde{P}_{X,Y,Z}$. 
\begin{lemma}
$\Lambda_\mo^{\prime \prime}(\lambda) >0$ for all $\lambda \in \bbR$. \eot
\label{lem:irreg-lem2} 
\end{lemma}

\begin{IEEEproof}
One can check that 
\begin{equation}
\Lambda_\mo^{\prime}(\lambda) = \mE_{\tilde{Q}_{X,Y,Z}^\lambda}\left[ \log \frac{W(Y|X)}{f^\ast(Y)}\right], \quad \Lambda_\mo^{\prime \prime}(\lambda) = \mV_{\tilde{Q}_{X,Y,Z}^\lambda}\left[ \log \frac{W(Y|X)}{f^\ast(Y)}\right]. 
\label{lem:irreg-lem2-pf1}
\end{equation}
For contradiction, assume there exists $\lambda \in \bbR$ with $\Lambda_\mo^{\prime \prime}(\lambda)=0$. We have
\begin{align}
\left[ \exists \, \lambda \in \bbR \textrm{ with } \Lambda_\mo^{\prime \prime}(\lambda) = 0 \right] \Longleftrightarrow & \left[ \log\frac{W(y|x)}{f^\ast(y)} = \Lambda_\mo^\prime(\lambda), \, \forall  \, (x,y,z) \in \tilde{\cS}_Q \right] \nonumber \\
  \Longrightarrow & \left[ \log\frac{W(y|x)}{f^\ast(y)} = \Lambda_\mo^\prime(\lambda), \, \forall  \, (x,y) \in \cS_Q \right]. \label{lem:irreg-lem2-pf2}
\end{align}
Using exactly the same arguments as in the proof of Lemma~\ref{lem:irreg-lem1}, one can show that \eqref{lem:irreg-lem2-pf2} contradicts item (i) of Lemma~\ref{lem:prelim}.  
\end{IEEEproof}

From item (i) of Lemma~\ref{lem:diff-new} and \eqref{eq:irreg-Lambda0-Lambda1}, we deduce that 
\begin{equation}
\Lambda_\mo^\prime\left(\frac{1-\rho^\ast}{1+\rho^\ast}\right) = - D_\mo.
\label{eq:irreg-bound-part2-2}
\end{equation}

Lemma~\ref{lem:irreg-lem2} and \eqref{eq:irreg-bound-part2-2} enable us to apply Lemma~\ref{lem:EA-new} in Appendix~\ref{app-rv-EA} to obtain 
\begin{equation}
\tilde{P}_{X,Y,Z}^N\left\{ \frac{1}{N}\sum_{n=1}^N \log\frac{W(Y_n|X_n)}{f^\ast(Y_N)} \geq - D_\mo \right\} \leq e^{-N \Lambda_\mo^\ast(-D_\mo)}\frac{1}{\sqrt{N}}\left\{ \frac{\tilde{m}_3}{\Lambda_\mo^{\prime \prime}(\tilde{\eta})^{3/2}} + \frac{1}{\sqrt{2 \pi \Lambda_\mo^{\prime \prime}(\tilde{\eta})}\tilde{\eta}}\right\}, \label{eq:irreg-bound-part2-3}
\end{equation}
where $\tilde{\eta} \eqdef \frac{1-\rho^\ast}{1+\rho^\ast}$, $\tilde{m}_3 \eqdef \mE_{\tilde{Q}_{X,Y,Z}^{\tilde{\eta}}}\left[ \left| \log \frac{W(Y|X)}{f^\ast(Y)} - \Lambda_\mo^\prime(\tilde{\eta})\right|^3\right]$ with $\tilde{Q}_{X,Y,Z}^{\tilde{\eta}}$ as defined in \eqref{eq:irreg-tilted-Q}, and 
\begin{equation}
\Lambda^\ast_\mo(-D_\mo) = \sup_{\lambda \in \bbR}\left\{ -D_\mo \lambda - \Lambda_\mo(\lambda) \right\}. 
\label{eq:irreg-bound-part2-4}
\end{equation}

Since $\Lambda_\mo(\cdot)$ is convex, \eqref{eq:irreg-bound-part2-2} and \eqref{eq:irreg-bound-part2-4} imply that
\begin{align}
\Lambda^\ast_\mo(-D_\mo) & = -\tilde{\eta}D_\mo - \Lambda_\mo\left( \tilde{\eta}\right) \nonumber \\
 & =  -\tilde{\eta}D_\mo - \Lambda_1([\tilde{\eta}, 1/(1+\rho^\ast)]^T), \label{eq:irreg-bound-part2-5}
\end{align}
where \eqref{eq:irreg-bound-part2-5} follows from \eqref{eq:irreg-Lambda0-Lambda1}. Item (ii) of Lemma~\ref{lem:diff-new} yields
\begin{equation}
\Lambda_1([\tilde{\eta}, 1/(1+\rho^\ast)]^T) = - \log P_{X,Y,Z}\left\{ \tilde{\cS}_Q\right\} + 2\Lambda\left( \frac{\rho^\ast}{1+\rho^\ast}\right). \label{eq:singular-new-1}
\end{equation}

Equations \eqref{eq:irreg-bound-part2-5} and \eqref{eq:singular-new-1} imply that 
\begin{align}
\Lambda_{\mo}^{\ast}(-D_\mo) & = \log P_{X,Y,Z}\left\{\tilde{\cS}_{Q}\right\} + \left[\left(\frac{\rho^\ast}{1+\rho^\ast} \right)\Lambda^\prime\left( \frac{\rho^\ast}{1+\rho^\ast}\right) - \Lambda\left( \frac{\rho^\ast}{1+\rho^\ast}\right)\right] \nonumber \\
 & \quad -\left[ \frac{1}{1+\rho^\ast}\Lambda^\prime\left( \frac{\rho^\ast}{1+\rho^\ast}\right) + \Lambda\left( \frac{\rho^\ast}{1+\rho^\ast}\right)\right] \nonumber \\
 & = \log P_{X,Y,Z}\left\{\tilde{\cS}_{Q}\right\} + \mE_r(R,Q) -\left[ \frac{1}{1+\rho^\ast}\Lambda^\prime\left( \frac{\rho^\ast}{1+\rho^\ast}\right) + \Lambda\left( \frac{\rho^\ast}{1+\rho^\ast}\right)\right] \label{eq:singular-new-2} \\
 & = \log P_{X,Y,Z}\left\{\tilde{\cS}_{Q}\right\} + \mE_r(R,Q) + R, \label{eq:singular-new-3}
\end{align}
where \eqref{eq:singular-new-2} follows from \eqref{eq:irreg-bound-part1-exp1} and \eqref{eq:irreg-bound-part1-exp1.75}, and \eqref{eq:singular-new-3} follows since
\[
-R =  \frac{1}{1+\rho^\ast}\Lambda^\prime\left( \frac{\rho^\ast}{1+\rho^\ast}\right) + \Lambda\left( \frac{\rho^\ast}{1+\rho^\ast}\right),
\]
which is \eqref{eq:lem5-3} in Appendix~\ref{app-prelim}. 

Equations \eqref{eq:irreg-bound-part2-1}, \eqref{eq:irreg-bound-part2-3} and \eqref{eq:singular-new-3} imply that
\[
\beta_N \leq e^{-N\left( \mE_\mr(R,Q) + R \right)}\frac{1}{\sqrt{N}}\left\{ \frac{\tilde{m}_3}{\Lambda_\mo^{\prime \prime}(\tilde{\eta})^{3/2}} + \frac{1}{\sqrt{2 \pi \Lambda^{\prime \prime}_{\mo}(\tilde{\eta})}\tilde{\eta}}\right\},
\]
which, in turn, implies that 
\begin{align}
(\lceil e^{NR} \rceil - 1) P_{X,Y,Z}^{N}\left\{ \frac{1}{N}\sum_{n=1}^{N}\log \frac{f^{\ast}(Y_{n})}{W(Y_{n}|X_{n})} \leq D_{\mo}, \frac{1}{N}\sum_{n=1}^{N} \log \frac{W(Y_{n}|X_{n})}{W(Y_{n}|Z_{n})}  \leq 0 \right\}  \leq \nonumber \\
\frac{e^{-N \mE_{\mr}(R,Q)}}{\sqrt{N}} \left\{ \frac{\tilde{m}_3}{\Lambda_\mo^{\prime \prime}(\tilde{\eta})^{3/2}} + \frac{1}{\sqrt{2 \pi \Lambda_\mo^{\prime \prime}(\tilde{\eta})}\tilde{\eta}}\right\} . \label{eq:irreg-bound-part2-final}
\end{align}
Plugging \eqref{eq:irreg-bound-part1} and \eqref{eq:irreg-bound-part2-final} into \eqref{eq:irreg-bound-1} implies \eqref{eq:thrm1.ii}. 

The proof of \eqref{eq:thrm1.ii-1} follows from the well-known expurgation idea (e.g., \cite[pg.~140]{gallager68}) and is included for completeness. To this end, generate a random code with $2\lceil e^{NR} \rceil$ codewords using $Q$ as specified in the beginning of this section. Using exactly the same arguments leading to the proof of \eqref{eq:thrm1.ii}, one can verify that
\begin{align}
\bar{\mP}_{\me}\left(Q,N,R + \frac{\log 2}{N}\right) & \leq \frac{e^{-N \mE_{\mr}(R,Q)}}{\sqrt{N}}\left\{ \frac{m_3}{\Lambda^{\prime \prime}(\eta)^{3/2}} + \frac{1}{\sqrt{2 \pi \Lambda^{\prime \prime}(\eta)}\eta} \right\} \nonumber \\
& \quad + \frac{e^{-N \mE_{\mr}(R,Q)}}{\sqrt{N}} \left\{ \frac{2\tilde{m}_3}{\Lambda_\mo^{\prime \prime}(\tilde{\eta})^{3/2}} + \frac{2}{\sqrt{2 \pi \Lambda_\mo^{\prime \prime}(\tilde{\eta})}\tilde{\eta}}\right\}\left\{ 1 + \frac{e^{-NR}}{2}\right\}.
\label{eq:singular-new-4} 
\end{align}
Clearly, \eqref{eq:singular-new-4} guarantees the existence of a code, say $(\tilde{f}, \tilde{\varphi})$, with blocklength $N$, $2 \lceil e^{NR}\rceil$ messages, and average error probability upper bounded by the right side of \eqref{eq:singular-new-4}. Now, if we throw out the worst (in terms of the corresponding conditional error probability) half of the codewords of this code, the resulting expurgated code, say $(f, \varphi)$, becomes an $(N,R)$ code with $\mP_{\me}(f, \varphi)$ not exceeding twice the right side of \eqref{eq:singular-new-4}, which, in turn, implies \eqref{eq:thrm1.ii-1}, which was to be shown.

\subsection{Proof of item (ii) of Theorem~\ref{thrm:thrm1}}
\label{ssec:proof-thrm1-i}
Assume the pair $(Q,W)$ is nonsingular. Let $\{ \epsilon_N \}_{N \geq 1}$ be such that $\epsilon_N = \frac{\log \sqrt{N}}{N}$ for all $N \in \bbZ^+$ and $R_N \eqdef R - \epsilon_N$. Consider a sufficiently large $N$ such that $R_N > R_\mcr(Q)$. For notational convenience, let 
\begin{equation}
\rho^\ast \eqdef - \left. \frac{\partial \mE_\mr(r, Q)}{\partial r}\right|_{r = R}, \quad  \rho^\ast_N \eqdef - \left. \frac{\partial \mE_\mr(r, Q)}{\partial r}\right|_{r = R_N}. 
\label{eq:nonsingular-rho}
\end{equation}
Let $f^\ast, \Lambda(\cdot)$ and $D_\mo$ denote the quantities defined in \eqref{eq:f-star}, \eqref{eq:lambda-N} and \eqref{eq:Do-N}, respectively, by choosing $\rho = \rho^\ast$. Similarly, let $f^\ast_N$, $\Lambda_N(\cdot)$ and $D_\mo(N)$ denote the quantities defined in \eqref{eq:f-star}, \eqref{eq:lambda-N} and \eqref{eq:Do-N}, respectively, by choosing $\rho = \rho^\ast_N$. Let $\cD_N$ denote the set defined in \eqref{eq:cDo-N}. Using these choices, \eqref{eq:thrm1-pf-step1-1.0} reads
\begin{align}
\bar{\mP}_{\me}(Q,N,R) & \leq P_{X,Y}^{N}\left\{ \cD_{N} \right\} \nonumber \\
& \quad + (\lceil e^{NR} \rceil- 1)P_{X,Y,Z}^{N}\left\{ \frac{1}{N}\sum_{n=1}^{N}\log\frac{f_N^\ast(Y_{n})}{W(Y_{n}|X_{n})}\leq D_{\mo}(N), \frac{1}{N}\sum_{n=1}^{N}\log\frac{W(Y_{n}|X_{n})}{W(Y_{n}|Z_{n})} \leq 0 \right\}. \label{eq:thrm1-pf-step1-1.0-new}
\end{align}

In order to conclude the proof, we must upper bound the two terms on the right side of \eqref{eq:thrm1-pf-step1-1.0-new}. We begin with the first term. 

Let $\eta_N \eqdef \frac{\rho^\ast_N}{1+\rho_N^\ast}$ and $\eta \eqdef \frac{\rho^\ast}{1+\rho^\ast}$. Item (iv) of Lemma~\ref{lem:prelim} ensures that $\rho_{(\cdot)}^\ast(Q)$ is continuous over $(R_\mcr(Q), \mI(Q;W))$ and hence, we have
\begin{align}
\lim_{N \rightarrow \infty} \rho_N^\ast & = \rho^\ast. \label{eq:thrm1-pf-rho-ast} \\
\lim_{N \rightarrow \infty} \eta_N & = \eta. \label{eq:limit-eta} \\
\lim_{N \rightarrow \infty} f^\ast_N(y) & = f^\ast(y) .\label{eq:thrm1-pf-f-ast} \\
\lim_{N \rightarrow \infty}\tilde{P}_{X,Y}^{\eta_N,\rho_N^\ast} & = \tilde{P}_{X,Y}^{\eta, \rho^\ast}.\label{eq:limiting-tilted-dist}
\end{align}

\begin{lemma}
Fix an arbitrary $\rho \in [0,1]$. For any $\lambda \in \bbR$, we have $\Lambda_\rho^{\prime \prime}(\lambda) \in \bbR^+$. \eot
\label{lem:lem-pos-variance}
\end{lemma}

\begin{IEEEproof}
Via elementary calculation, one can check that 
\begin{equation}
\Lambda_\rho^\prime(\lambda) = \mE_{\tilde{P}_{X,Y}^{\lambda,\rho}}\left[ \log \frac{f_\rho(Y)}{W(Y|X)}\right], \quad \Lambda_\rho^{\prime \prime}(\lambda) = \mV_{\tilde{P}_{X,Y}^{\lambda,\rho}}\left[ \log \frac{f_\rho(Y)}{W(Y|X)}\right] \geq 0, \label{eq:lem-pos-variance-pf2}
\end{equation}
where $\tilde{P}_{X,Y}^{\lambda, \rho}$ is defined in \eqref{eq:lem-pos-variance-pf1}. The inequality in \eqref{eq:lem-pos-variance-pf2} ensures that it suffices to prove $\Lambda_\rho^{\prime \prime}(\cdot) \neq 0$. For contradiction, assume this is not the case. Then, 
\begin{align}
\left[ \exists \, \lambda \in \bbR \textrm{ s.t. } \Lambda_\rho^{\prime \prime}(\lambda) = 0 \right]  \Longleftrightarrow  & \left[ \log \frac{f_\rho(Y)}{W(Y|X)} = \Lambda^\prime_\rho(\lambda), \, \forall (x,y) \in \cS_Q \right] \nonumber \\
 \Longrightarrow & \left[ W(y|x) = W(y|z), \, \forall (x,y,z) \in \tilde{\cS}_Q \right]. \label{eq:lem-pos-variance-pf3}
\end{align}
The right side of \eqref{eq:lem-pos-variance-pf3} is equivalent to saying that the pair $(Q,W)$ is singular, which is a contradiction. Hence, we conclude that $\Lambda_\rho^{\prime \prime}(\lambda) >0$. 
\end{IEEEproof}

Lemma~\ref{lem:lem-pos-variance} ensures that $\Lambda^{\prime \prime}(\cdot), \Lambda_N^{\prime \prime}(\cdot) \in \bbR^+$, thus we can apply Lemma~\ref{lem:diff-new} in Appendix~\ref{app-rv-EA} to obtain\footnote{In the conference version of this work, the second term in the braces of \eqref{eq:1-upper-bound1} is incorrectly written as $\frac{1}{\sqrt{2 \pi}\eta_N}$ \cite[Eq. (29)]{altug-allerton12}. The correct form is $\frac{1}{\sqrt{2 \pi \Lambda_N^{\prime \prime}(\eta_N)}\eta_N}$, as given in \eqref{eq:1-upper-bound1}.}
\begin{equation}
P_{X,Y}^N\left\{ \cD_N \right\} \leq e^{-N \Lambda_N^\ast(D_\mo(N))}\frac{1}{\sqrt{N}}\left\{ \frac{m_{3,N}}{\Lambda_N^{\prime \prime}(\eta_N)^{3/2}} + \frac{1}{\sqrt{2 \pi \Lambda_N^{\prime \prime}(\eta_N)}\eta_N}\right\}, \label{eq:1-upper-bound1}
\end{equation}
where $m_{3,N} \eqdef \mE_{\tilde{P}_{X,Y}^{\eta_N, \rho_N^\ast}}\left[ \left| \log\frac{f_N^\ast(Y)}{W(Y|X)} - \Lambda_N^\prime(\eta_N) \right|^3\right]$ and $\Lambda_N^\ast(D_\mo(N))$ is the Fenchel-Legendre transform of $\Lambda_N(\cdot)$ at $D_{\mo}(N)$.

Since $\Lambda_N(\cdot)$ is convex, one can verify that 
\begin{equation}
\Lambda_N^\ast(D_\mo(N)) = \eta_N \Lambda_N^\prime(\eta_N) - \Lambda_N(\eta_N). 
\label{eq:fenchel-2}
\end{equation}

Lemma~\ref{lem:lem3} and \eqref{eq:lem5-2} in Appendix~\ref{app-prelim} imply that 
\begin{equation}
\mE_\mr(R_N, Q)= \eta_N \Lambda^\prime_N(\eta_N) - \Lambda_N(\eta_N). \label{eq:nonsingular-new1}
\end{equation}
By substituting \eqref{eq:nonsingular-new1} into \eqref{eq:fenchel-2}, we deduce that
\begin{equation}
\Lambda_N^\ast(D_\mo(N)) = \mE_\mr(R_N, Q). 
\label{eq:fenchel-3}
\end{equation}

By using \eqref{eq:thrm1-pf-rho-ast}--\eqref{eq:lem-pos-variance-pf2}, along with the continuity of $|\cdot|^3$ and $(\cdot)^2$, and the fact that $\cX$ and $\cY$ are finite sets, we conclude that 
\begin{align}
\lim_{N \rightarrow \infty} \Lambda_N^{\prime \prime}(\eta_N) & = \Lambda^{\prime \prime}(\eta) \label{eq:limit-Lambda-double-prime}, \\
\lim_{N \rightarrow \infty} m_{3,N} & = m_3 \eqdef \mE_{\tilde{P}_{X,Y}^{\eta, \rho^\ast}}\left[ \left| \log\frac{f^\ast(Y)}{W(Y|X)} - \Lambda^\prime(\eta) \right|^3\right]. \label{eq:limit-m3}
\end{align}

Due to \eqref{eq:limit-eta}, \eqref{eq:limit-Lambda-double-prime} and \eqref{eq:limit-m3}, one can choose a sufficiently large $N$ with 
\begin{equation}
 \frac{m_{3,N}}{\Lambda_N^{\prime \prime}(\eta_N)^{3/2}} + \frac{1}{\sqrt{2 \pi \Lambda_N^{\prime \prime}(\eta_N)}\eta_N} \leq  2\left(\frac{m_{3}}{\Lambda^{\prime \prime}(\eta)^{3/2}} + \frac{1}{\sqrt{2 \pi \Lambda^{\prime \prime}(\eta)}\eta}\right). \label{eq:constant-limit}
\end{equation}
By substituting \eqref{eq:fenchel-3} and \eqref{eq:constant-limit} into \eqref{eq:1-upper-bound1}, we deduce that
\begin{equation}
P_{X,Y}^N\left\{ \cD_N \right\} \leq \frac{2}{\sqrt{N}}\left( \frac{m_{3}}{\Lambda^{\prime \prime}(\eta)^{3/2}} + \frac{1}{\sqrt{2 \pi \Lambda^{\prime \prime}(\eta)}\eta}\right)e^{-N \mE_\mr(R_N, Q)}. \label{eq:1-upper-bound2}
\end{equation}

Next, we upper bound the second term on the right side of \eqref{eq:thrm1-pf-step1-1.0}. 

To begin with, note that for any $(x,y,z)$ with $Q(x)W(y|x)Q(z) >0$, if $(x,y,z) \notin \tilde{\cS}_{Q}$, then $ \log\frac{W(y|x)}{W(y|z)} = \infty$, which, in turn, implies that 
\begin{align}
\alpha_{N} & \eqdef P_{X,Y,Z}^{N}\left\{ \frac{1}{N}\sum_{n=1}^{N}\log\frac{f_{N}^{\ast}(Y_{n})}{W(Y_{n}|X_{n})} \leq D_{\mo}(N), \frac{1}{N}\sum_{n=1}^{N}\log\frac{W(Y_{n}|X_{n})}{W(Y_{n}|Z_{n})} \leq 0 \right\} \nonumber \\
 & = P_{X,Y,Z}^{N}\left\{ \tilde{\cS}_{Q}^{N}\right\}\tilde{\alpha}_{N}, \label{eq:thrm1-pf-step3-1}
\end{align}
where, in \eqref{eq:thrm1-pf-step3-1} we define
\begin{equation}
\tilde{\alpha}_{N} \eqdef \tilde{P}_{X,Y,Z}^{N}\left\{ \frac{1}{N}\sum_{n=1}^{N}\log\frac{W(Y_{n}|X_{n})}{f_{N}^{\ast}(Y_{n})} \geq -D_{\mo}(N), \frac{1}{N}\sum_{n=1}^{N}\log\frac{W(Y_{n}|Z_{n})}{W(Y_{n}|X_{n})} \geq 0 \right\}. \label{eq:talphaN}
\end{equation}

Given any $\bv \in \bbR^{2}$ let $\Lambda_{1,N}(\bv)$ and $\Lambda_{1}(\bv)$ denote $\Lambda_{1, \rho_N^\ast}(\bv)$ and $\Lambda_{1, \rho^\ast}(\bv)$, respectively, where $\Lambda_{1,\rho}(\bv)$ is defined in \eqref{eq:lambda1}. Further, define 
\begin{equation}
\bv^{\ast}(N) \eqdef \left[ \frac{1-\rho_{N}^{\ast}}{1+\rho_{N}^{\ast}}, \frac{1}{1+\rho_{N}^{\ast}}\right]^{T}, \quad \bv^{\ast} \eqdef \left[ \frac{1-\rho^{\ast}}{1+\rho^{\ast}}, \frac{1}{1+\rho^{\ast}}\right]^{T}.
\label{eq:s-t-ast}
\end{equation}
Note that $\bv^{\ast}_{1}, \bv^{\ast}_{1}(N) \in (0,1)$ and $\bv_{2}^{\ast}, \bv^{\ast}_{2}(N) \in (1/2,1)$. Also, by using \eqref{eq:thrm1-pf-rho-ast}--\eqref{eq:thrm1-pf-f-ast}, one can verify that 
\begin{align}
\lim_{N \rightarrow \infty} \bv^\ast(N) & = \bv^\ast, \label{eq:v-limit} \\
\lim_{N \rightarrow \infty} \Lambda_{1,N}(\bv^\ast(N)) & =  \Lambda_{1}(\bv^\ast). \label{eq:Lambda1-limit}
\end{align}

Given any $\rho \in [0,1]$ and $\bv \in \bbR^2$, define 
\begin{equation}
\tilde{Q}_{X,Y,Z}^{\bv, \rho}(x,y,z) \eqdef 
\begin{cases}
\frac{\tilde{P}_{X,Y,Z}(x,y,z)W(y|x)^{\bv_{1} - \bv_{2}} f_{\rho}(y)^{-\bv_{1}} W(y|z)^{ \bv_{2}}}{\sum_{(a,b,c) \in \tilde{S}_{Q}}\tilde{P}_{X,Y,Z}(a,b,c)W(b|a)^{\bv_{1} - \bv_{2}} f_{\rho}(b)^{-\bv_{1}} W(b|c)^{ \bv_{2}}} & \textrm{ if } (x,y,z) \in \tilde{S}_{Q} \\
 0 & \textrm{ else.}
\end{cases} 
\label{eq:tilted-Q-N}
\end{equation}
Note that $\tilde{Q}_{X,Y,Z}^{\bv, \rho}$ is a well-defined probability measure and equivalent to $\tilde{P}_{X,Y,Z}$. For notational convenience, let $\tilde{Q}_{X,Y,Z}^{\bv^{\ast}(N)}$ and $\tilde{Q}_{X,Y,Z}^{\bv^{\ast}}$ denote $\tilde{Q}_{X,Y,Z}^{\bv^\ast(N), \rho^\ast_N}$ and $\tilde{Q}_{X,Y,Z}^{\bv^\ast, \rho^\ast}$, respectively. 

From \eqref{eq:thrm1-pf-f-ast}, \eqref{eq:v-limit} and \eqref{eq:tilted-Q-N}, we deduce that 
\begin{equation}
\lim_{N \rightarrow \infty} \tilde{Q}_{X,Y,Z}^{\bv^\ast(N)} = \tilde{Q}_{X,Y,Z}^{\bv^\ast}. \label{eq:tilted-Q-limit}
\end{equation}

In the remaining part of the proof, we need the following result whose validity heavily depends on the nonsingularity of the pair $(Q,W)$. 
\begin{lemma}
Fix an arbitrary $r \in \left(R_\mcr(Q), \mI(Q;W) \right)$. Let $\rho \eqdef -\left. \frac{\partial \mE_\mr(a,Q)}{\partial a}\right|_{a = r} \in (0,1)$ and $\tilde{\bv} \eqdef \left[ \frac{1-\rho}{1+\rho}, \frac{1}{1+\rho}\right]^T$. We have 
\begin{equation}
\det\left( \textrm{cov}_{\tilde{Q}_{X,Y,Z}^{\tilde{\bv}, \rho}, \rho}\left( \left[ \log\frac{W(Y|X)}{f_\rho(Y)}, \log \frac{W(Y|Z)}{W(Y|X)}\right]^T \right) \right) > 0. \label{eq:tilted-cov}  
\end{equation}
\eot
\label{lem:lem-step5}
\end{lemma}
\begin{IEEEproof}
The proof is given in Appendix~\ref{app-lem-step5}.
\end{IEEEproof}

Define
\begin{align}
& \bb(N) \eqdef [-D_{\mo}(N), 0]^{T}, \quad \bb \eqdef [-D_\mo, 0]^T, \quad \cB(N) \eqdef [-D_{\mo}(N), \infty) \times [0, \infty). \label{eq:cB}\\
& \Lambda_{1,N}^{\ast}(\bd) \eqdef \sup_{\bv \, \in \, \bbR^{2}}\left\{ \langle  \bv, \bd \rangle - \Lambda_{1,N}(\bv)\right\}, \label{eq:Lambda1-fenchel}
\end{align}
for any $\mathbf{d} \in \bbR^{2}$.

For notational convenience, let 
\[
\bS_N \eqdef \textrm{cov}_{\tilde{Q}_{X,Y,Z}^{\bv^\ast(N)}}\left( \left[ \log\frac{W(Y|X)}{f_N^\ast(Y)}, \log \frac{W(Y|Z)}{W(Y|X)}\right]^T \right), \bS \eqdef \textrm{cov}_{\tilde{Q}_{X,Y,Z}^{\bv^\ast}}\left( \left[ \log\frac{W(Y|X)}{f^\ast(Y)}, \log \frac{W(Y|Z)}{W(Y|X)}\right]^T \right), 
\]
and note that \eqref{eq:tilted-cov} ensures that $\lambda_{\textrm{min}}(\bS_N), \lambda_{\textrm{min}}(\bS) \in \bbR^+$, where $\lambda_{\textrm{min}}(\bS_N)$ (resp. $\lambda_{\textrm{min}}(\bS)$) denotes the minimum eigenvalue of $\bS_N$ (resp. $\bS$). 

\begin{lemma}
For all sufficiently large $N$ that depends on $Q$, $W$ and $R$, 
\begin{equation}
\tilde{\alpha}_N \leq e^{-N \Lambda_{1,N}^\ast(\bb(N))} \frac{c}{2\lambda_{\min}(\mathbf{\Sigma}_N)N}\left( k(R,W,Q)^2 + \frac{2}{\bv_1^\ast(N)^2} + \frac{2}{\bv_2^\ast(N)^2}\right),
\label{eq:vector-EA}
\end{equation}
where $c \in \bbR^+$ is a universal constant and $k(R,W,Q) \in \bbR^+$ is a constant that depends on $R, W$ and $Q$. \eot
\label{lem:vector-EA}
\end{lemma}
\begin{IEEEproof}
The proof is given in Appendix~\ref{app-vector-EA}. 
\end{IEEEproof}

\begin{remark}
Although we state Lemma~\ref{lem:vector-EA} in the context of our setup, its extension to general i.i.d.\ random vectors satisfying the usual regularity conditions required for strong large deviations results is evident. Moreover, this result gives a more general upper bound than the existing vector exact asymptotics results of Chaganty and Sethuraman~\cite{chaganty-sethuraman96} and Petrovskii~\cite{petrovskii96}. In particular, \cite{chaganty-sethuraman96} and \cite{petrovskii96} handle strongly non-lattice random vectors\footnote{A random vector is strongly non-lattice if the magnitude of its characteristic function is bounded away from $1$ everywhere, except the origin.} and lattice random vectors\footnote{A random vector is lattice if it only takes values on a lattice.}, respectively. Unlike the case of scalars, however, for random vectors these two cases are not exhaustive, and we are not aware of a result that gives an upper bound of $O(1/N)$ without such a restriction. \eor
\end{remark}

As a direct consequence of the fact that the eigenvalues of a real square matrix depend continuously upon its entries (e.g., \cite[App.~D]{horn-johnson85}), we have the following
\begin{lemma}
For all sufficiently large $N$, 
\begin{equation}
\lambda_{\textrm{min}}(\bS_{N}) \geq \frac{\lambda_{\textrm{min}}(\bS)}{2\sqrt{2}}.
\label{eq:min-eigenvalue}
\end{equation}
\eot
\label{lem:min-eigenvalue}
\end{lemma}

Further, due to \eqref{eq:v-limit} and $\bv_{1}^{\ast}, \bv_{2}^{\ast} \in \bbR^{+}$, we have 
\begin{equation}
 \frac{1}{\bv_1^\ast(N)^2} + \frac{1}{\bv_2^\ast(N)^2} \leq \frac{2}{(\bv_1^\ast)^2} + \frac{2}{(\bv_2^\ast)^2},
\label{eq:2nd-term-bound}
\end{equation}
for all sufficiently large $N$.

Plugging \eqref{eq:min-eigenvalue} and \eqref{eq:2nd-term-bound} into \eqref{eq:vector-EA}, we finally deduce that 
\begin{equation}
\tilde{\alpha}_{N}   \leq e^{-N \Lambda_{1,N}^\ast(\bb(N))} \frac{4\sqrt{2}c}{\lambda_{\min}(\mathbf{\Sigma})N}\left( \frac{k(R,W,Q)^2}{4} + \frac{1}{(\bv_1^\ast)^2} + \frac{1}{(\bv_2^\ast)^2}\right),
\label{eq:2nd-term-bound-final}
\end{equation}
for all sufficiently large $N$. 

Next, we deal with the exponent in \eqref{eq:2nd-term-bound-final}. First of all, owing to the convexity of $\Lambda_{1,N}(\cdot)$ and item (i) of Lemma~\ref{lem:diff-new}, one can show that 
\begin{equation}
\Lambda_{1,N}^{\ast}(\bb(N)) = -\bv_1^{\ast}(N)D_{\mo}(N) - \Lambda_{1,N}(\bv^\ast(N)).
\label{eq:Lambda1-fenchel-aB}
\end{equation}

Item (ii) of Lemma~\ref{lem:diff-new} and \eqref{eq:Lambda1-fenchel-aB}, along with the definitions of $D_\mo(N)$ and $\bv^\ast(N)$, imply that  
\begin{align}
\Lambda_{1,N}^{\ast}(\bb(N)) & =\log P_{X,Y,Z}\left\{\tilde{\cS}_{Q}\right\} + \left[\left(\frac{\rho_N^\ast}{1+\rho_N^\ast} \right)\Lambda_N^\prime\left( \frac{\rho_N^\ast}{1+\rho_N^\ast}\right) - \Lambda_N\left( \frac{\rho_N^\ast}{1+\rho_N^\ast}\right)\right] \nonumber \\
 & \quad -\left[ \frac{1}{1+\rho_N^\ast}\Lambda_N^\prime\left( \frac{\rho_N^\ast}{1+\rho_N^\ast}\right) + \Lambda_N\left( \frac{\rho_N^\ast}{1+\rho_N^\ast}\right)\right] \nonumber \\
 & =\log P_{X,Y,Z}\left\{\tilde{\cS}_{Q}\right\} + \mE_r(R_N,Q) -\left[ \frac{1}{1+\rho_N^\ast}\Lambda_N^\prime\left( \frac{\rho_N^\ast}{1+\rho_N^\ast}\right) + \Lambda_N\left( \frac{\rho_N^\ast}{1+\rho_N^\ast}\right)\right] \label{eq:thrm1-pf-exponent1} \\
 & =\log P_{X,Y,Z}\left\{\tilde{\cS}_{Q}\right\} + \mE_r(R_N,Q) + R_N, \label{eq:thrm1-pf-exponent2}
\end{align}
where \eqref{eq:thrm1-pf-exponent1} follows from \eqref{eq:fenchel-2} and \eqref{eq:fenchel-3}, and \eqref{eq:thrm1-pf-exponent2} follows from \eqref{eq:lem5-3} in Appendix~\ref{app-prelim}. 

By using \eqref{eq:2nd-term-bound-final}, \eqref{eq:thrm1-pf-exponent2} and the fact that $\epsilon_N = \frac{\log N}{2N}$, we have 
\begin{equation}
\tilde{\alpha}_N \leq P_{X,Y,Z}\left\{ \tilde{\cS}_Q \right\}^{-N} \frac{4\sqrt{2}c}{\lambda_{\min}(\mathbf{\Sigma})\sqrt{N}}\left( \frac{k(R,W,Q)^2}{4} + \frac{1}{(\bv_1^\ast)^2} + \frac{1}{(\bv_2^\ast)^2}\right) e^{-N\left( \mE_\mr(R_N,Q) + R\right)}.
\label{eq:2nd-term-bound-final.01}
\end{equation}

Since $P_{X,Y,Z}^N\left\{\tilde{\cS}_Q^N\right\} = P_{X,Y,Z}\left\{ \tilde{\cS}_Q\right\}^N$,  \eqref{eq:thrm1-pf-step3-1} and \eqref{eq:2nd-term-bound-final.01} imply that 
\begin{equation}
\alpha_N \leq \frac{4\sqrt{2}c}{\lambda_{\min}(\mathbf{\Sigma})\sqrt{N}}\left( \frac{k(R,W,Q)^2}{4} + \frac{1}{(\bv_1^\ast)^2} + \frac{1}{(\bv_2^\ast)^2}\right) e^{-N\left( \mE_\mr(R_N,Q) + R\right)}. 
\label{eq:2nd-term-bound-final.1}
\end{equation}

Equation \eqref{eq:2nd-term-bound-final.1} finally implies that 
\begin{align}
(\lceil e^{NR} \rceil - 1) P_{X,Y,Z}^{N}\left\{ \frac{1}{N}\sum_{n=1}^{N}\log\frac{f^\ast_N(Y_{n})}{W(Y_{n}|X_{n})}\leq D_{\mo}(N), \frac{1}{N}\sum_{n=1}^{N}\log\frac{W(Y_{n}|X_{n})}{W(Y_{n}|Z_{n})}\leq 0 \right\}  = (\lceil e^{NR} \rceil - 1) \alpha_N \nonumber \\
  \leq \frac{4\sqrt{2}c}{\lambda_{\min}(\mathbf{\Sigma})\sqrt{N}}\left( \frac{k(R,W,Q)^2}{4} + \frac{1}{(\bv_1^\ast)^2} + \frac{1}{(\bv_2^\ast)^2}\right) e^{-N\mE_\mr(R_N,Q)}.\label{eq:2nd-term-bound-final.2}
\end{align}

Plugging \eqref{eq:1-upper-bound2} and \eqref{eq:2nd-term-bound-final.2} into \eqref{eq:thrm1-pf-step1-1.0} yields, 
\begin{align}
\bar{\mP}_{\me,m}(Q,N,R) & \leq \frac{2}{\sqrt{N}}\left\{ \frac{m_{3}}{\Lambda^{\prime \prime}(\eta)^{3/2}} + \frac{1}{\sqrt{2 \pi \Lambda^{\prime \prime}(\eta)}\eta}\right\}e^{-N \mE_\mr(R_N, Q)} \nonumber \\
& \quad + \frac{4\sqrt{2}c}{\lambda_{\min}(\mathbf{\Sigma})\sqrt{N}}\left( \frac{k(R,W,Q)^2}{4} + \frac{1}{(\bv_1^\ast)^2} + \frac{1}{(\bv_2^\ast)^2}\right) e^{-N\mE_\mr(R_N,Q)}. \label{eq:final-bound}
\end{align}

Evident convexity of $\mE_{\mr}(\cdot, Q)$, along with its continuous differentiability over $[R_N, R]$, which is ensured by item (iv) of Lemma~\ref{lem:prelim}, enables us to deduce that (e.g., \cite[eq.~(3.2)]{boyd-vand2004})
\begin{equation}
\mE_\mr(R_N,Q) \geq \mE_\mr(R,Q) - \frac{\log N}{2N}\left.\frac{\d \mE_{\mr}(r,Q)}{\d r}\right|_{r=R}. 
\label{eq:final-bound-exponent}
\end{equation}

Equations \eqref{eq:final-bound} and \eqref{eq:final-bound-exponent} imply \eqref{eq:thrm1.i}. 

The proof of \eqref{eq:thrm1.i-1} follows from the same arguments leading to the proof of \eqref{eq:thrm1.ii-1}, which are given below for completeness. First, generate a random code with $2 \lceil e^{NR} \rceil$ codewords using $Q$ as specified in the beginning of this section. Using exactly the same arguments leading to the proof of \eqref{eq:thrm1.i}, one can verify that
\begin{align}
\bar{\mP}_{\me,m}\left(Q,N,R + \frac{\log 2}{N}\right) & \leq \frac{2}{\sqrt{N}}\left\{ \frac{m_{3}}{\Lambda^{\prime \prime}(\eta)^{3/2}} + \frac{1}{\sqrt{2 \pi \Lambda^{\prime \prime}(\eta)}\eta}\right\}e^{-N \mE_\mr(R_N, Q)} \nonumber \\
& \quad + \frac{8\sqrt{2}c}{\lambda_{\min}(\mathbf{\Sigma})\sqrt{N}}\left( \frac{k(R,W,Q)^2}{4} + \frac{1}{(\bv_1^\ast)^2} + \frac{1}{(\bv_2^\ast)^2}\right)\left( 1 + \frac{e^{-NR}}{2}\right) e^{-N\mE_\mr(R_N,Q)}. \label{eq:final-bound-1}
\end{align}
Clearly, \eqref{eq:final-bound-1} guarantees the existence of a code, say $(\tilde{f}, \tilde{\varphi})$, with blocklength $N$, $2\lceil e^{NR}\rceil$ messages and average error probability upper bounded by the right side of \eqref{eq:final-bound-1}. Now, if we throw out the worst (in terms of the corresponding conditional error probability) half of the codewords of this code, the resulting expurgated code, say $(f, \varphi)$, becomes an $(N,R)$ code with $\mP_\me\left( f, \varphi \right)$ not exceeding twice the right side of \eqref{eq:final-bound-1}, which, in turn, implies \eqref{eq:thrm1.i-1}, which was to be shown.


\section{Proof of Theorem~\ref{thrm:thrm3}}
\label{app-thrm3}
Let $W \in \cP(\cY|\cX)$ be arbitrary with $R_{\mcr}<C$. 
\begin{itemize}
\item[(i)] For any $R \in \bbR_+$, we write $\mE_\mr(R)$ as 
\begin{equation}
\mE_\mr(R) = \max_{(\rho, Q) \in [0,1] \times \cP(\cX)}\psi_R(\rho, Q), \label{eq:thrm3-pf1}
\end{equation}
where $\psi_R(\rho,Q) \eqdef -\rho R + \mE_\mo(\rho, Q)$. For any $(\rho, Q) \in [0,1] \times \cP(\cX)$, $\psi_{(\cdot)}(\rho, Q)$ is a linear function, and hence convex and continuous over $(R_\mcr,C)$. Further, given any $R \in (R_\mcr , C)$, $\psi_R(\cdot, \cdot)$ is continuous over $[0,1] \times \cP(\cX)$ (e.g., \cite[Lemma~2.1]{altug12-a}), and evidently $[0,1] \times \cP(\cX) \subset \bbR^{|\cX|+1}$ is a compact set. 

Now, fix an arbitrary $R \in (R_\mcr, C)$, and note that due to the observations in the previous paragraph, we can apply a well-known result from convex analysis (e.g., \cite[Theorem~2.87]{Ruszczynski2006}), namely that the subdifferential of the maximum function satisfies 
\begin{equation}
\partial \mE_\mr(R) = \textnormal{conv}\left( \left\{ - \rho^\ast: (\rho^\ast, Q^\ast) \in [0,1] \times \cP(\cX) \textnormal{ achieves the maximum in } \eqref{eq:thrm3-pf1} \right\} \right).
\label{eq:thrm3-pf2}
\end{equation}
Due the fact that $R_{\mcr} < R < C$, one can verify that for any $(\rho^\ast, Q^\ast) \in [0,1] \times \cP(\cX)$ that achieves the maximum in \eqref{eq:thrm3-pf1}, $\rho^\ast \in (0,1)$. Hence, items (iii) and (iv) of Lemma~\ref{lem:prelim}, along with \eqref{eq:thrm3-pf2}, imply that 
\begin{equation*}
\partial \mE_\mr(R)  = \textnormal{conv}\left( \left\{ \left.\frac{\partial \mE_\mr(a,Q^\ast)}{\partial a}\right|_{a = R}  : \mE_\mr(R,Q^\ast) = \mE_\mr(R) \right\}\right),
\end{equation*}
which is \eqref{eq:thrm3-i}. 

\item[(ii)] Since $\mE_\mr(\cdot)$ is a real-valued, convex function over $[R_\mcr, C]$, $\partial \mE_\mr(R)$, i.e., the subdifferential of $\mE_\mr(\cdot)$ at $R$, is a nonempty, convex and compact set (e.g.,~\cite[Theorem~2.74]{Ruszczynski2006}), for all $R \in (R_\mcr, C)$. Thus, $\rho_R^\ast$ is well-defined. Equation \eqref{eq:thrm3-ii} is an evident consequence of item (ii) of Theorem~\ref{thrm:thrm1} by invoking it with the $Q \in \cP(\cX)$ whose existence is assumed in the statement of the theorem. 

\item[(iii)] Consider any positive channel $W$. First, we note that for any $Q \in \cP(\cX)$, if the pair $(Q,W)$ is singular, then there exists $\delta_y \in \bbR^+$ such that $W(y|x) = \delta_y$, for all $y \in \cY$ and $x \in \cX$ with $Q(x) >0$. Now, consider any $R \in (R_\mcr, C)$ and $Q \in \cP(\cX)$ with $\mE_\mr(R, Q) = \mE_\mr(R)$. For contradiction, assume that the pair $(Q,W)$ is singular. Due to the observation at the beginning of this item, along with the positivity of the channel, one can verify that $\mE_\mo(\rho, Q) = -\log\sum_y \delta_y$, for all $\rho \in \bbR_+$, which contradicts item (i) of Lemma~\ref{lem:prelim}. Hence, we conclude that the pair $(Q,W)$ should be nonsingular. This, in light of the definition of $\rho^\ast_R$ and item (i) of this lemma, suffices to conclude the proof.   
\end{itemize}

\section{Proof of Theorem~\ref{thrm:thrm2}}
\label{sec:proof-thrm2}
As pointed out in the statement of the theorem, item (ii) is due to Gallager and hence we only prove item (i). Let $W \in \cP(\cY|\cX)$ with $C >0$ and $R \leq R_{\mcr}$ be arbitrary. Consider some $Q \in \cP(\cX)$ with $\mE_{\mo}(1,Q) = \max_{P \in \cP(\cX)}\mE_{\mo}(1,P)$, such that the pair $(Q,W)$ is singular. For this $(Q,W)$ pair, define 
\begin{align*}
P_{X,Y,Z}(x,y,z) \eqdef Q(x)W(y|x)Q(z), \, \forall (x,y,z) \in \cX \times \cY \times \cX, \\
\tilde{P}_{X,Y,Z}(x,y,z) \eqdef 
\begin{cases}
\frac{P_{X,Y,Z}(x,y,z)}{P_{X,Y,Z}\left\{ \tilde{\cS}_{Q}\right\}} & \textrm{ if } (x,y,z) \in \tilde{\cS}_{Q}, \\
0 & \textrm{ else}.
\end{cases}
\end{align*}
similar to \eqref{eq:P_XYZ} and \eqref{eq:tP_XYZ}. Let $\tilde{\cS}_{Q}$ and $\cX_{y}$ be as in \eqref{eq:tSQ} and \eqref{eq:Xy}, respectively, for this choice of $(Q,W)$. 

First, we show that 
\begin{equation}
\log P_{X,Y,Z}\left\{ \tilde{\cS}_{Q}\right\} = - \mE_{\mo}(1,Q). 
\label{eq:thrm2-ii-pf1}
\end{equation}
To see this, note that  
\begin{align}
\log P_{X,Y,Z}\left\{ \tilde{\cS}_{Q}\right\} & = \log \sum_{(x,y,z) \in \tilde{\cS}_{Q}} Q(x)W(y|x)Q(z) \nonumber \\
 & = \log \sum_{(x,y,z) \in \tilde{\cS}_{Q}} Q(x)W(y|x)^{1/2}Q(z)W(y|z)^{1/2} \label{eq:thrm2-ii-pf2} \\
 & = \log \sum_{y} \left[ \sum_{x \in \cS(Q) \cap \cX_{y}}Q(x)W(y|x)^{1/2}\right]\left[ \sum_{z \in \cS(Q) \cap \cX_{y}}Q(z)W(y|z)^{1/2}\right] \nonumber \\
 & = -\mE_{\mo}(1,Q), \nonumber 
\end{align}
where \eqref{eq:thrm2-ii-pf2} follows from the singularity of $(Q,W)$. 

Further, for any message $m$
\begin{align}
\bar{\mP}_{\me,m}(Q, N, R) & \leq \left(\lceil e^{NR} \rceil -1 \right)  P_{X,Y,Z}^{N}\left\{ \frac{1}{N}\sum_{n=1}^{N}\log \frac{W(Y_{n}|X_{n})}{W(Y_{n}|Z_{n})} \leq 0 \right\} \nonumber \\
& = \left(\lceil e^{NR} \rceil -1 \right) P_{X,Y,Z}\left\{ \tilde{\cS}_{Q}\right\}^{N}\tilde{P}_{X,Y,Z}^{N}\left\{ \frac{1}{N}\sum_{n=1}^{N}\log \frac{W(Y_{n}|X_{n})}{W(Y_{n}|Z_{n})} \leq 0 \right\} \label{eq:thrm2-ii-pf3} \\
& = \left(\lceil e^{NR} \rceil -1\right) P_{X,Y,Z}\left\{ \tilde{\cS}_{Q}\right\}^{N} \label{eq:thrm2-ii-pf4} \\
& \leq e^{-N(-R + \mE_{\mo}(1,Q))} \label{eq:thrm2-ii-pf5} \\
& =  e^{-N\mE_{\mr}(R)}, \label{eq:thrm2-ii-pf6} 
\end{align}
where \eqref{eq:thrm2-ii-pf3} follows from the fact that for any $(x,y,z)$ with $Q(x)W(y|x)Q(z)>0$, if $(x,y,z) \notin \tilde{\cS}_{Q}$, then $\log \frac{W(y|x)}{W(y|z)} = \infty$, \eqref{eq:thrm2-ii-pf4} follows from the singularity of $(Q,W)$, \eqref{eq:thrm2-ii-pf5} follows from \eqref{eq:thrm2-ii-pf1} and \eqref{eq:thrm2-ii-pf6} is true because of the choice of $Q \in \cP(\cX)$ and the fact that $R \leq R_{\mcr}$ (e.g., \cite[pg.~245]{gallager73}). Hence, the upper bound of \eqref{eq:thrm2-1} follows.

In order to establish the lower bound of \eqref{eq:thrm2-1}, one can use Gallager's arguments \cite[pg.~245-246]{gallager73}, and hence we conclude the proof.


\appendices

\section{Proof of Lemma~\ref{lem:prelim}}
\label{app:lem-prelim-pf}

Throughout this section, fix an arbitrary $W \in \cP(\cY|\cX)$ such that $R_\mcr < C$, and $Q \in \cP(\cX)$ such that $\mE_\mr(R,Q) >0$ for some $R > R_\infty$.
\begin{itemize}
\item[(i)] Since $\mE_\mr(R,Q)  \in \bbR^+$, one can see that $R \in (0, \mI(Q;W))$. This observation enables us to invoke \cite[Theorem~5.6.3]{gallager68}, which, in turn, ensures that 
\begin{equation}
\frac{\partial^2 \mE_{\mo}(\rho, Q)}{\partial \rho^2} \leq 0,
\label{eq:lem-prelim-pf-1}
\end{equation}
for all $\rho \in \bbR_+$. Moreover, \cite[Theorem~5.6.3]{gallager68} also guarantees that if \eqref{eq:lem-prelim-pf-1} holds with equality for some $\rho \in \bbR_+$, then the same should be true for all $\rho \in \bbR_+$. To draw a contradiction, assume \eqref{eq:lem-prelim-pf-1} holds with equality for some $\rho \in \bbR_+$, which, in turn, implies that $\frac{\partial \mE_{\mo}(\rho, Q)}{\partial \rho} = \mI(Q;W) $ for all $\rho \in \bbR_+$, due to \cite[Eq.~(5.6.25)]{gallager68}. Since $\mE_{\mo}(0, Q) =0$, we have 
\begin{equation}
\mE_\mo(\rho, Q) = \rho \mI(Q;W). 
\label{eq:lem-prelim-pf-2}
\end{equation}
To conclude the proof, consider 
\begin{equation} 
\mE_{\mSP}(R,Q) \eqdef \sup_{\rho \geq 0 } \left\{ -\rho R + \mE_{\mo}(\rho, Q)\right\},
\label{eq:lem-prelim-pf-3}
\end{equation}
and notice that substituting \eqref{eq:lem-prelim-pf-2} into \eqref{eq:lem-prelim-pf-3} yields $\mE_\mSP(R,Q) = \infty$, which contradicts $R > R_\infty$. 

\item[(ii)] Fix any $R_\infty < r \leq \mI(Q;W)$. Equation \eqref{eq:lem-prelim-1} is a direct consequence of the fact that $R_\infty < r$. Further, since $\left.\frac{ \partial \mE_{\mo}(\rho,Q)}{\partial \rho}\right|_{\rho=0} = \mI(Q;W)$ (e.g., \cite[Eq.~(5.6.25)]{gallager68}), item (i) of this lemma suffices to conclude the proof of this item. 

\item[(iii)] The assertion follows from \eqref{eq:lem-prelim-2}, along with the fact that $\left.\frac{ \partial \mE_{\mo}(\rho,Q)}{\partial \rho}\right|_{\rho=0} = \mI(Q;W)$ and item (i) of this lemma.  

\item[(iv)] Fix some $r \in \left(  \left. \frac{\partial \mE_{\mo}(\rho, Q) }{\partial \rho}\right|_{\rho =1}  , \mI(Q;W)\right)$, and consider 
\begin{equation}
\mE_\mr(r, Q) = \max_{\rho \in [0,1]}\left\{ -\rho r + \mE_\mo(\rho, Q)\right\}.
\label{eq:lem-prelim-pf-4}
\end{equation}
Using the the characterization of the subdifferential of the maximum function (e.g.,~\cite[Theorem~2.87]{Ruszczynski2006}), we have 
\begin{equation}
\partial \mE_\mr(\cdot, Q)(a) = \textnormal{conv}\left( \{ -\rho^\ast  : \mE_\mr(r,Q) = -\rho^\ast  r + \mE_{\mo}(\rho^\ast , Q)   \}\right).
\label{eq:lem-prelim-pf-5}
\end{equation}
Items (ii) and (iii) of this lemma ensures that \eqref{eq:lem-prelim-pf-4} has a unique maximizer, which is $\rho_r^\ast(Q)$. Therefore, \eqref{eq:lem-prelim-pf-5} reduces to 
\begin{equation*}
\partial \mE_\mr(\cdot, Q)(r) = \{ - \rho_r^\ast(Q)\},
\end{equation*}
which, in turn, implies \eqref{eq:lem-prelim-3}, and hence we conclude the proof. 
\end{itemize}


\section{Auxiliary Results}
\label{app-prelim}
This section contains some auxiliary results that will be used in the proof Theorem~\ref{thrm:thrm1}. Throughout the  section, fix an arbitrary $W \in \cP(\cY|\cX)$ with $R_\mcr < C$, and $Q \in \cP(\cX)$ with $\mE_\mr(R,Q) >0$ for some $R > R_\infty$. Fix some\footnote{The non-emptiness of the following interval is ensured by item (i) of Lemma~\ref{lem:prelim}.} $r \in \left(  -\left. \frac{\partial \mE_{\mo}(\rho, Q)}{\partial \rho}\right|_{\rho =1} , \mI(Q;W) \right)$. Let $\rho_r^\ast(Q) \eqdef - \left. \frac{\partial \mE_{\mr}(a, Q)}{\partial a}\right|_{a = r}$, which is well-defined due to \eqref{eq:lem-prelim-3}, and note that $\rho_r^\ast(Q) \in (0,1)$, because of item (iii) of Lemma~\ref{lem:prelim}. 

\begin{definition}
\begin{enumerate}
\item[(i)] For any $y \in \cY$ and $\rho \in \bbR_+$, 
\begin{equation}
P_Y^\rho(y) \eqdef \frac{\left[ \sum_{x \in \cX}Q(x)W(y|x)^{1/(1+\rho)}\right]^{1+\rho}}{\sum_{b \in \cY}\left[ \sum_{a \in \cX}Q(a)W(b|a)^{1/(1+\rho)}\right]^{1+\rho}}. \label{eq:P_Y}
\end{equation}
Observe that $P_Y^\rho$ is a well-defined probability measure on $\cY$, for any $\rho \in \bbR_{+}$. For notational convenience, we define $f^\ast_r \eqdef P_Y^{\rho_r^\ast(Q)}$.
\item[(ii)] For any $\rho \in \bbR_+$, 
\begin{equation}
P_{X|Y}^\rho(x|y) \eqdef 
\begin{cases}
\frac{Q(x)W(y|x)^{1/(1+\rho)}}{\sum_{a \in \cX}Q(a)W(y|a)^{1/(1+\rho)}} & \textrm{ if } y \in \cS(P_Y^\rho), \\
0 & \textrm{else}.
\end{cases}
\label{eq:P_X|Y}
\end{equation}
Note that $P^{\rho}_{X|Y}$ is a well-defined conditional probability measure for all $\rho \in \bbR_{+}$.  

\item[(iii)] For any $(x,y) \in \cX \times \cY$ and $\rho \in \bbR_+$
\begin{equation}
P_{X,Y}^\rho(x,y) \eqdef  P_{X|Y}^\rho(x|y)P_Y^\rho(y). 
\label{eq:P_XY}
\end{equation}
For notational convenience, we let $P^0_{X,Y}(x,y) =: P_{X,Y}(x,y) = Q(x)W(y|x)$, for any $(x,y) \in \cX \times \cY$.

\item[(iv)] 
\begin{equation}
\mE_\mF(r,Q) \eqdef \mD\left( P_{X,Y}^{\rho_r^\ast(Q)}||Q \times W\right) = \sum_{(x,y) \in \cX \times \cY}P_{X,Y}^{\rho_r^\ast(Q)}(x,y) \log \frac{P_{X,Y}^{\rho_r^\ast(Q)}(x,y)}{Q(x)W(y|x)}.
\label{eq:EF}
\end{equation}

\item[(v)] For any $\lambda \in \bbR$, 
\begin{equation}
\Lambda_r(\lambda) \eqdef \log \mE_{P_{X,Y}}\left[ e^{\lambda \log\frac{f^\ast_r(Y)}{W(Y|X)}}\right]. 
\label{eq:Lambda}
\end{equation}
\end{enumerate}
\label{def:def1}
\eod
\end{definition}

\begin{lemma}
\begin{equation}
\frac{\d \mE_\mo(\rho,Q)}{\d \rho} = \sum_{(x,y) \in \cX \times \cY}P_{X,Y}^\rho(x,y)\log \frac{P_{X|Y}^\rho(x|y)}{Q(x)},
\label{eq:lem2}
\end{equation}
for all $\rho \in \bbR_+$. \eot
\label{lem:lem2}
\end{lemma}

\begin{IEEEproof}
Define $h_y(\rho, Q) \eqdef \sum_{x \in \cX}Q(x)W(y|x)^{1/(1+\rho)}$ and $g_y(\rho, Q) \eqdef h_y(\rho, Q)^{1+\rho}$. From the definition of $\mE_\mo(\cdot, \cdot)$, i.e., \eqref{eq:Eo},  
\begin{equation}
\frac{\d \mE_{\mo}(\rho, Q)}{\d \rho} = -\frac{\sum_{y \in \cY} \frac{\d g_y(\rho, Q)}{\d \rho}}{\sum_{b \in \cY} g_b(\rho, Q)}. 
\label{eq:lem2-pf1}
\end{equation}
Note that if $\cS(Q) \cap \cX_y = \emptyset$, then $h_y(\rho, Q) = g_y(\rho, Q) = 0$ for all $\rho \in \bbR_+$. Also, observe that there exists $y \in \cY$, such that $\cS(Q) \cap \cX_y \neq \emptyset$. Further, one can check that provided that $\cS(Q) \cap \cX_y \neq \emptyset$,
\begin{align}
\frac{\d h_y(\rho, Q)}{\d \rho} & = -\frac{1}{(1+\rho)^2}\sum_{x \in \cX} Q(x)W(y|x)^{1/(1+\rho)}\log W(y|x), \label{eq:lem2-pf2} \\
\frac{\d g_y(\rho, Q)}{\d \rho} & = g_y(\rho, Q)\left[ (1+\rho)\frac{\frac{\d h_y(\rho, Q)}{\d \rho}}{h_y(\rho, Q)} + \log h_y(\rho, Q)\right]. \label{eq:lem2-pf3}
\end{align}

Equations \eqref{eq:lem2-pf1} and \eqref{eq:lem2-pf3} imply that 
\begin{align}
\frac{\d \mE_\mo(\rho, Q)}{\d \rho} & = -\sum_{y:\cX_y \cap \cS(Q) \neq \emptyset} \frac{g_y(\rho, Q)}{\sum_{b \in \cY}g_b(\rho, Q)}\left[(1+\rho)\frac{\frac{\d h_y(\rho, Q)}{\d \rho}}{h_y(\rho, Q)} + \log h_y(\rho, Q) \right] \nonumber \\
& = -\sum_{y:\cX_y \cap \cS(Q) \neq \emptyset} P_Y^\rho(y)\left[(1+\rho)\frac{\frac{\d h_y(\rho, Q)}{\d \rho}}{h_y(\rho, Q)} + \log h_y(\rho, Q) \right], \label{eq:lem2-pf4}
\end{align}
where \eqref{eq:lem2-pf4} follows from the definition of $P_Y^\rho$, i.e., \eqref{eq:P_Y}. Consider any $y$ with $\cX_y \cap \cS(Q) \neq \emptyset$. We have
\begin{align}
(1+\rho)\frac{\frac{\d h_y(\rho, Q)}{\d \rho}}{h_y(\rho, Q)} + \log h_y(\rho, Q) & =  \log \sum_{z \in \cX} Q(z)W(y|z)^{1/(1+\rho)} +\sum_{x \in \cX} \frac{Q(x)W(y|x)^{1/(1+\rho)}}{\sum_{a \in \cX} Q(a)W(y|a)^{1/(1+\rho)}} \log \frac{1}{W(y|x)^{\frac{1}{1+\rho}}} \label{eq:lem2-pf5}\\
 & = \sum_{x \in \cX}P_{X|Y}^{\rho}(x|y)\log\frac{1}{W(y|x)^{\frac{1}{1+\rho}}}+\sum_{x \in \cX}P_{X|Y}^{\rho}(x|y)\log \sum_{z \in \cX} Q(z)W(y|z)^{\frac{1}{1+\rho}} \label{eq:lem2-pf6}\\
 & = \sum_{x \in \cX}P_{X|Y}^{\rho}(x|y) \log\frac{Q(x)}{P_{X|Y}^{\rho}(x|y)}, \label{eq:lem2-pf7}
\end{align}
where \eqref{eq:lem2-pf5} follows from \eqref{eq:lem2-pf2}, \eqref{eq:lem2-pf6} and \eqref{eq:lem2-pf7} follow from the definition of $P_{X|Y}^\rho$, i.e., \eqref{eq:P_X|Y}. Plugging \eqref{eq:lem2-pf7} into \eqref{eq:lem2-pf4} and remembering the definition of $P_{X,Y}^\rho$, i.e., \eqref{eq:P_XY}, we conclude that \eqref{eq:lem2} holds. 
\end{IEEEproof}

\begin{lemma}
\begin{equation}
\mE_{\mF}(r,Q) = \mE_{\mr}(r,Q). 
\label{eq:lem3}
\end{equation}
\label{lem:lem3}
\eot
\end{lemma}

\begin{IEEEproof}
Observe that owing to the definitions of $P_{X|Y}^\rho$ and $P_{X,Y}^\rho$, i.e., \eqref{eq:P_X|Y} and \eqref{eq:P_XY}, along with the definition of $\mE_F(r, Q)$, i.e., \eqref{eq:EF}, we have
\begin{equation}
\mE_{\mF}(r,Q) = \sum_{(x,y) \in \cS_{Q,W}}P_{X,Y}^{\rho_{r}^{\ast}(Q)}(x,y)\log\frac{P_{Y}^{\rho_{r}^{\ast}(Q)}(y)}{W(y|x)^{\frac{\rho_{r}^{\ast}(Q)}{1+\rho_{r}^{\ast}(Q)}}\left[\sum_{a \in \cX}Q(a) W(y|a)^{\frac{1}{1+\rho_{r}^{\ast}(Q)}}\right]}. \label{eq:lem3-pf1}
\end{equation}

Moreover, 
\begin{align}
\mE_{\mr}(r,Q) & = - r \rho_{r}^{\ast}(Q) + \mE_{\mo}(\rho_{r}^{\ast}(Q), Q) \label{eq:lem3-pf2} \\
 & = - \rho_{r}^{\ast}(Q) \sum_{(x,y) \in \cS_{Q,W}}P_{X,Y}^{\rho_{r}^{\ast}(Q)}(x,y)\log\frac{P_{X|Y}^{\rho_{r}^{\ast}(Q)}(x|y)}{Q(x)}+\mE_{\mo}(\rho_{r}^{\ast}(Q), Q) \label{eq:lem3-pf3} \\
 & = \sum_{(x,y) \in \cS_{Q,W}}P_{X,Y}^{\rho_{r}^{\ast}(Q)}(x,y) \log\frac{\left(\sum_{z \in \cX}Q(z)W(y|z)^{\frac{1}{1+\rho_{r}^{\ast}(Q)}}\right)^{\rho_{r}^{\ast}(Q)}}{W(y|x)^{\frac{\rho_{r}^{\ast}(Q)}{1+\rho_{r}^{\ast}(Q)}}\left[ \sum_{b \in \cY}\left( \sum_{a \in \cX}Q(a)W(b|a)^{\frac{1}{1+\rho_{r}^{\ast}(Q)}}\right)^{1+\rho_{r}^{\ast}(Q)}\right]} \label{eq:lem3-pf4} \\
& =  \sum_{(x,y) \in \cS_{Q,W}}P_{X,Y}^{\rho_{r}^{\ast}(Q)}(x,y)\log\frac{P_{Y}^{\rho_{r}^{\ast}(Q)}(y)}{W(y|x)^{\frac{\rho_{r}^{\ast}(Q)}{1+\rho_{r}^{\ast}(Q)}}\left[\sum_{a \in \cX}Q(a)W(y|a)^{\frac{1}{1+\rho_{r}^{\ast}(Q)}}\right]}, \label{eq:lem3-pf5}
\end{align}
where \eqref{eq:lem3-pf2} follows from item (i) of Lemma~\ref{lem:prelim} and \eqref{eq:lem-prelim-1},  \eqref{eq:lem3-pf3} follows from \eqref{eq:lem2}, \eqref{eq:lem3-pf4} follows from the definition of $\mE_\mo(\rho, Q)$, i.e., \eqref{eq:Eo}, and the definition of $P_{X|Y}^\rho$, i.e., \eqref{eq:P_X|Y}, and \eqref{eq:lem3-pf5} follows from the definition of $P_Y^\rho$, i.e., \eqref{eq:P_Y}. Equations \eqref{eq:lem3-pf1} and \eqref{eq:lem3-pf5} together imply \eqref{eq:lem3}. 
\end{IEEEproof}

\begin{lemma}
\begin{align}
\Lambda_r\left( \frac{\rho_{r}^{\ast}(Q)}{1+\rho_{r}^{\ast}(Q)}\right) & = \frac{1}{1+\rho_{r}^{\ast}(Q)}\log\sum_{y \in \cY} \left[ \sum_{x \in \cX} Q(x)W(y|x)^{\frac{1}{1+\rho_{r}^{\ast}(Q)}}\right]^{1+\rho_{r}^{\ast}(Q)}. \label{eq:lem5-0} \\
 \Lambda^{\prime}_r\left( \frac{\rho_{r}^{\ast}(Q)}{1+\rho_{r}^{\ast}(Q)}\right)  & = \sum_{(x,y) \in \cS_{Q,W}}P_{X,Y}^{\rho_{r}^{\ast}(Q)}(x,y)\log\frac{f^\ast_r(y)}{W(y|x)}. \label{eq:lem5-1}\\
 \mE_{\mF}(r,Q)  & = \frac{\rho_{r}^{\ast}(Q)}{1+\rho_{r}^{\ast}(Q)}\Lambda_r^{\prime}\left(\frac{\rho_{r}^{\ast}(Q)}{1+\rho_{r}^{\ast}(Q)}\right)-\Lambda_r\left( \frac{\rho_{r}^{\ast}(Q)}{1+\rho_{r}^{\ast}(Q)}\right). \label{eq:lem5-2} \\
  r & = -\frac{1}{1+\rho_{r}^{\ast}(Q)}\Lambda_r^{\prime}\left( \frac{\rho_{r}^{\ast}(Q)}{1+\rho_{r}^{\ast}(Q)}\right) - \Lambda_r\left( \frac{\rho_{r}^{\ast}(Q)}{1+\rho_{r}^{\ast}(Q)}\right). \label{eq:lem5-3}
\end{align}
\label{lem:lem5}
\eot
\end{lemma}

\begin{IEEEproof}
From the definition of $P_Y^\rho$, i.e., \eqref{eq:P_Y}, we have 
\begin{equation*}
\Lambda_r\left(\frac{\rho_{r}^{\ast}(Q)}{1+\rho_{r}^{\ast}(Q)}\right) = \log \sum_{(x,y) \in \cS_{Q,W}}Q(x)W(y|x)^\frac{1}{1+\rho_{r}^{\ast}(Q)}\left[ \frac{\left(\sum_{z \in \cX} Q(z)W(y|z)^{\frac{1}{1+\rho_{r}^{\ast}(Q)}}\right)^{1+\rho_{r}^{\ast}(Q)}}{\sum_{b \in \cY}\left( \sum_{a \in \cX} Q(a)W(b|a)^{\frac{1}{1+\rho_{r}^{\ast}(Q)}}\right)^{1+\rho_{r}^{\ast}(Q)}}\right]^{\frac{\rho_{r}^{\ast}(Q)}{1+\rho_{r}^{\ast}(Q)}}, \label{eq:lem5-pf01}
\end{equation*}
which, in turn, implies that 
\begin{equation*}
\Lambda_r\left(\frac{\rho_{r}^{\ast}(Q)}{1+\rho_{r}^{\ast}(Q)}\right) = \frac{1}{1+\rho_{r}^{\ast}(Q)}\log \sum_{y \in \cY}\left( \sum_{x \in \cX} Q(x)W(y|x)^{\frac{1}{1+\rho_{r}^{\ast}(Q)}}\right)^{1+ \rho_{r}^{\ast}(Q)}.
\end{equation*}

Next, one can check that
\begin{equation}
\Lambda_r^\prime\left( \frac{\rho_r^\ast(Q)}{1+\rho_r^\ast(Q)}\right) = \sum_{(x,y) \in \cS_{Q,W}}\frac{Q(x)W(y|x)^\frac{1}{1+\rho_r^\ast(Q)}f_r^\ast(y)^{\frac{\rho_r^\ast(Q)}{1+\rho_r^\ast(Q)}}}{\sum_{(a,b) \in \cS_{Q,W}}Q(a)W(b|a)^{\frac{1}{1+\rho_r^\ast(Q)}}f_r^\ast(b)^{\frac{\rho_r^\ast(Q)}{1+\rho_r^\ast(Q)}}}\log\frac{f_r^\ast(y)}{W(y|x)}. \label{eq:lem5-pf1}
\end{equation}
By the definition of $P_Y^\rho$, i.e., \eqref{eq:P_Y}, for any $(x,y) \in \cS_{Q,W}$, we have
\begin{align}
\frac{Q(x)W(y|x)^{\frac{1}{1+\rho_r^\ast(Q)}}f_r^\ast(y)^{\frac{\rho_r^\ast(Q)}{1+\rho_r^\ast(Q)}}}{\sum_{(a,b) \in \cS_{Q,W}}Q(a)W(b|a)^{\frac{1}{1+\rho_r^\ast(Q)}}f_r^\ast(b)^{\frac{\rho_r^\ast(Q)}{1+\rho_r^\ast(Q)}}} & = 
\frac{Q(x)W(y|x)^{\frac{1}{1+\rho_r^\ast(Q)}}\left[ \sum_{z \in \cX} Q(z)W(y|z)^{\frac{1}{1+\rho_r^\ast(Q)}}\right]^{\rho_r^\ast(Q)}}{\sum_{(a,b)\in \cS_{Q,W}}Q(a)W(b|a)^{\frac{1}{1+\rho_r^\ast(Q)}}\left[ \sum_{c \in \cX} Q(c)W(b|c)^{\frac{1}{1+\rho_r^\ast(Q)}}\right]^{\rho_r^\ast(Q)}} \nonumber \\
& = \frac{Q(x)W(y|x)^{\frac{1}{1+\rho_r^\ast(Q)}}\left[ \sum_{z \in \cX} Q(z)W(y|z)^{\frac{1}{1+\rho_r^\ast(Q)}}\right]^{\rho_r^\ast(Q)}}{\sum_{b \in \cY}\left[ \sum_{a \in \cX} Q(a)W(b|a)^{\frac{1}{1+\rho_r^\ast(Q)}}\right]^{1+\rho_r^\ast(Q)}} \nonumber \\
& = P_{X|Y}^{\rho_r^\ast(Q)}(x|y)P_Y^{\rho_r^\ast(Q)}(y) \label{eq:lem5-pf2} \\
& = P_{X,Y}^{\rho_r^\ast(Q,W)}(x,y), \label{eq:lem5-pf3}
\end{align}
where \eqref{eq:lem5-pf2} follows from the definitions of $P_Y^\rho$ and $P_{X|Y}^\rho$, i.e., \eqref{eq:P_Y} and \eqref{eq:P_X|Y}, \eqref{eq:lem5-pf3} follows from the definition of $P_{X,Y}^\rho$, i.e., \eqref{eq:P_XY}. Plugging \eqref{eq:lem5-pf3} into \eqref{eq:lem5-pf1} implies \eqref{eq:lem5-1}. 

From the definition of $\mE_\mF(r,Q)$, i.e., \eqref{eq:EF}, and the definition of $P_Y^\rho$, i.e., \eqref{eq:P_Y}, we have 
\begin{align}
\mE_\mF(r,Q) & = \sum_{(x,y) \in \cS_{Q,W}}P_{X,Y}^{\rho_r^\ast(Q)}(x,y)\log\frac{P_Y^{\rho_r^\ast(Q)}(y)}{W(y|x)} + \sum_{(x,y) \in \cS_{Q,W}}P_{X,Y}^{\rho_r^\ast(Q)}(x,y)\log\frac{P_{X|Y}^{\rho_r^\ast(Q)}(x|y)}{Q(x)} \nonumber \\
 & = \Lambda^\prime\left(\frac{\rho_r^\ast(Q)}{1+\rho_r^\ast(Q)}\right) + \sum_{(x,y)\in \cS_{Q,W}}P_{X,Y}^{\rho_r^\ast(Q)}\log\frac{W(y|x)^{\frac{1}{1+\rho_r^\ast(Q)}}}{\sum_{z \in \cX} Q(z)W(y|z)^{\frac{1}{1+\rho_r^\ast(Q)}}} \label{eq:lem5-pf4} \\
 & = \frac{\rho_r^\ast(Q)}{1+\rho_r^\ast(Q)}\Lambda_r^\prime\left( \frac{\rho_r^\ast(Q)}{1+\rho_r^\ast(Q)}\right) + \sum_{(x,y) \in \cS_{Q,W}}P_{X,Y}^{\rho_r^\ast(Q)} \log \frac{f_r^\ast(y)^{1/(1+\rho_r^\ast(Q))}}{\sum_{z \in \cX}Q(z)W(y|z)^{\frac{1}{1+\rho_r^\ast(Q)}}}\label{eq:lem5-pf4.1} \\
 & =  \frac{\rho_r^\ast(Q)}{1+\rho_r^\ast(Q)}\Lambda_r^\prime\left( \frac{\rho_r^\ast(Q)}{1+\rho_r^\ast(Q)}\right)   + \sum_{(x,y) \in \cS_{Q,W}}P_{X,Y}^{\rho_r^\ast(Q)} \log \frac{1}{\left[\sum_{b}\left( \sum_{a}Q(a)W(b|a)^{\frac{1}{1+\rho_r^\ast(Q)}}\right)^{(1+\rho_r^\ast(Q))} \right]^{\frac{1}{1+\rho_r^\ast(Q)}}} \label{eq:lem5-pf4.2} \\
  & = \frac{\rho_r^\ast(Q)}{1+\rho_r^\ast(Q)}\Lambda_r^\prime\left( \frac{\rho_r^\ast(Q)}{1+\rho_r^\ast(Q)}\right) - \Lambda_r\left(\frac{\rho_r^\ast(Q)}{1+\rho_r^\ast(Q)}\right), \label{eq:lem5-pf5}
\end{align}
where \eqref{eq:lem5-pf4} and \eqref{eq:lem5-pf4.1} follow from \eqref{eq:lem5-1}, \eqref{eq:lem5-pf4.2} follows from the definition of $P_Y^\rho$, i.e., \eqref{eq:P_Y}, and \eqref{eq:lem5-pf5} follows from \eqref{eq:lem5-0}. 

Lastly, the fact that $\left.\frac{\partial \mE_\mo(\rho, Q)}{\partial \rho}\right|_{\rho = \rho_r^\ast(Q)}= r$, which is established in \eqref{eq:lem-prelim-2}, along with Lemma~\ref{lem:lem2}, implies that 
\begin{align}
r & = \sum_{(x,y) \in \cS_{Q,W}}P_{X,Y}^{\rho_{r}^{\ast}(Q)}(x,y) \log\frac{P_{X|Y}^{\rho_{r}^{\ast}(Q)}(x|y)}{Q(x)} \nonumber \\
 & = \sum_{(x,y) \in \cS_{Q,W}}P_{X,Y}^{\rho_{r}^{\ast}(Q)}(x,y)\log\frac{P_{X,Y}^{\rho_{r}^{\ast}(Q)}(x,y)}{Q(x)W(y|x)} + \sum_{(x,y) \in \cS_{Q,W}}P_{X,Y}^{\rho_{r}^{\ast}(Q)}(x,y)\log\frac{W(y|x)}{P_{Y}^{\rho_{r}^{\ast}(Q)}(y)} \nonumber \\
 & = \mE_{\mF}(r,Q) - \Lambda_r^{\prime}\left( \frac{\rho_{r}^{\ast}(Q)}{1+\rho_{r}^{\ast}(Q)}\right) \label{eq:lem5-pf6}\\
 & = -\frac{1}{1+\rho_{r}^{\ast}(Q)}\Lambda_r^{\prime}\left( \frac{\rho_{r}^{\ast}(Q)}{1+\rho_{r}^{\ast}(Q)}\right) - \Lambda_r\left( \frac{\rho_{r}^{\ast}(Q)}{1+\rho_{r}^{\ast}(Q)}\right), \label{eq:lem5-pf7}
 \end{align}
 where \eqref{eq:lem5-pf6} follows from the definition of $\mE_\mF(r,Q)$, i.e., \eqref{eq:EF}, \eqref{eq:lem2} and \eqref{eq:lem5-1}, and \eqref{eq:lem5-pf7} follows from \eqref{eq:lem5-2}. 
\end{IEEEproof}

\section{Proof of Lemma~\ref{lem:diff-new}}
\label{app-lem-diff-new-pf}
\begin{itemize}
\item[(i)] By elementary calculation, 
\begin{equation}
\frac{\d \Lambda_{1,\rho}(\bv_1,\bv_2)}{\d \bv_2} = \sum_{(x,y,z)\in \tilde{\cS}_{Q}}\frac{Q(x)W(y|x)^{1+\bv_1-\bv_2}f_\rho(y)^{-\bv_1}Q(z)W(y|z)^{\bv_2}}{\sum_{(a,b,c) \in \tilde{\cS}_{Q}}Q(a)W(b|a)^{1+\bv_1-\bv_2}f_\rho(b)^{-\bv_1}Q(c)W(b|c)^{\bv_2}}\log\frac{W(y|z)}{W(y|x)}, 
\label{eq:thrm1-pf-step5-2}
\end{equation}
and
\begin{equation}
\frac{\d \Lambda_{1,\rho}(\bv_1,\bv_2)}{\d \bv_1} = \sum_{(x,y,z)\in \tilde{\cS}_{Q}}\frac{Q(x)W(y|x)^{1+\bv_1-\bv_2}f_\rho(y)^{-\bv_1}Q(z)W(y|z)^{\bv_2}}{\sum_{(a,b,c) \in \tilde{\cS}_{Q}}Q(a)W(b|a)^{1+\bv_1-\bv_2}f_\rho(b)^{-\bv_1}Q(c)W(b|c)^{\bv_2}}\log\frac{W(y|x)}{f_\rho(y)}. 
\label{eq:thrm1-pf-step5-2.5} 
\end{equation} 


Evaluating the right side of \eqref{eq:thrm1-pf-step5-2} at $\tilde{\bv}$ yields\footnote{Note that the particular value of $\tilde{\bv}_2$ does not matter as long as one has $\tilde{\bv}_1 = -1+2\tilde{\bv}_2$.}
\begin{equation}
\left.\frac{\d \Lambda_{1,\rho}(\tilde{\bv}_1,\bv_2)}{\d \bv_2}\right|_{\bv_2 = \tilde{\bv}_2} = 0,
\label{eq:thrm1-pf-step5-3}
\end{equation}
owing to the symmetry of the resulting expression. 

Equation \eqref{eq:thrm1-pf-step5-2.5} further implies that 
\begin{equation}
\frac{\d \Lambda_{1,\rho}(\bv_1,\bv_2)}{\d \bv_1}  = \sum_{(x,y) \in \cS_{Q}}\frac{Q(x)W(y|x)^{1+\bv_1-\bv_2}f_\rho(y)^{-\bv_1}\left[ \sum_{z \in \cS(Q)\cap\cX_{y}}Q(z)W(y|z)^{\bv_2}\right]}{\sum_{(a,b)\in\cS_{Q}}Q(a)W(b|a)^{1+\bv_1-\bv_2}f_\rho(b)^{-\bv_1}\left[ \sum_{c \in \cS(Q)\cap\cX_{b}}Q(c)W(b|c)^{\bv_2}\right]}\log\frac{W(y|x)}{f_\rho(y)}. \label{eq:thrm1-pf-step5-4}
\end{equation}

Evaluating the right side of \eqref{eq:thrm1-pf-step5-4} at $\tilde{\bv}$ yields
\begin{align}
& \left. \frac{\d \Lambda_{1,\rho}(\bv_1,\tilde{\bv}_2)}{\d \bv_1}\right|_{\bv_1 = \tilde{\bv}_1}  \nonumber \\
 & =\sum_{(x,y) \in \cS_{Q}}\frac{Q(x)W(y|x)^{\tilde{\bv}_2}f_\rho(y)^{1-2\tilde{\bv}_2}\left[ \sum_{z \in \cS(Q)\cap\cX_{y}}Q(z)W(y|z)^{\tilde{\bv}_2}\right]}{\sum_{(a,b)\in\cS_{Q}}Q(a)W(b|a)^{\tilde{\bv}_2}f_\rho(b)^{1-2\tilde{\bv}_2}\left[ \sum_{c \in \cS(Q)\cap\cX_{b}}Q(c)W(b|c)^{\tilde{\bv}_2}\right]} \log\frac{W(y|x)}{f_\rho(y)}. \label{eq:thrm1-pf-step5-5}
\end{align}
Note that for any $y \in \cY$, such that $\cX_{y} \cap \cS(Q) \neq \emptyset$, we have
\begin{equation}
\left[ \left(\sum_{x}Q(x)W(y|x)^{1/(1+\rho)}\right)^{1+\rho}\right]^{-\tilde{\bv}_2}= \frac{1}{\sum_{x}Q(x)W(y|x)^{1/(1+\rho)}}. \label{eq:thrm1-pf-step5-6}
\end{equation}
By substituting \eqref{eq:thrm1-pf-step5-6} into \eqref{eq:thrm1-pf-step5-5}, along with the definition of $f_\rho$ and  \eqref{eq:lem5-1} in Appendix~\ref{app-prelim}, we conclude that
\begin{equation}
\left. \frac{\d \Lambda_{1,\rho}(\bv_1,\tilde{\bv}_2)}{\d \bv_1}\right|_{\bv_1 = \tilde{\bv}_1} = -\Lambda_\rho^{\prime}\left(\frac{\rho}{1+\rho}\right). 
\label{eq:thrm1-pf-step5-7}
\end{equation}
Equations \eqref{eq:thrm1-pf-step5-3} and \eqref{eq:thrm1-pf-step5-7} together imply \eqref{eq:tilted-mean2-1}, which was to be shown. 

\item[(ii)] Note that 
\begin{align}
\Lambda_{1,\rho}(\tilde{\bv}) & = \log \sum_{(x,y,z) \in \tilde{\cS}_{Q}}\tilde{P}_{X,Y,Z}(x,y,z)\left(\frac{W(y|x)}{f_\rho(y)}\right)^{\tilde{\bv}_1} \left( \frac{W(y|z)}{W(y|x)}\right)^{\tilde{\bv}_2} \nonumber \\
 & = -\log P_{X,Y,Z}\left\{ \tilde{\cS}_{Q}\right\} + \nu_{\tilde{\bv}}, \label{eq:thrm1-pf-step5-9.1}
\end{align}
where we define
\begin{equation}
\nu_{\tilde{\bv}} \eqdef \log \sum_{(x,y,z) \in \tilde{\cS}_{Q}}Q(x)W(y|x)^{\tilde{\bv}_2}Q(z)W(y|z)^{\tilde{\bv}_2}f_\rho(y)^{-\tilde{\bv}_1}. 
\label{eq:thrm1-pf-step5-10.1}
\end{equation}

Observe that for any $y \in \cY$ such that $ \cX_{y} \cap \cS(Q) \neq \emptyset$, we have 
\begin{equation}
f_\rho(y)^{-\tilde{\bv}_1}  = \frac{f_\rho(y)^{\rho/(1+\rho)}}{\sum_{x}Q(x)W(y|x)^{1/(1+\rho)}}\left[ \sum_{b}\left(\sum_{a}Q(a)W(b|a)^{1/(1+\rho)} \right)^{1+\rho}\right]^{1/(1+\rho)},
\label{eq:thrm1-pf-step5-11.1}
\end{equation}
owing to the definitions of $f_\rho$ and $\tilde{\bv}$. Rearranging \eqref{eq:thrm1-pf-step5-11.1} gives
\begin{equation}
\sum_{z}Q(z)W(y|z)^{1/(1+\rho)}f_\rho(y)^{-\tilde{\bv}_1} = f_\rho(y)^{\rho/(1+\rho)}\left[ \sum_{b}\left(\sum_{a}Q(a)W(b|a)^{1/(1+\rho)} \right)^{1+\rho}\right]^{1/(1+\rho)},
\label{eq:thrm1-pf-step5-12.1}
\end{equation}
provided that $y \in \cY$ satisfies $\cX_{y} \cap \cS(Q) \neq \emptyset$. By substituting \eqref{eq:thrm1-pf-step5-12.1} into \eqref{eq:thrm1-pf-step5-10.1} and noting the definition of $\tilde{\bv}$, we deduce that 
\begin{align}
\nu_{\tilde{\bv}} & = \log \sum_{(x,y) \in \cS_{Q}}Q(x)W(y|x)^{1/(1+\rho)}f_\rho(y)^{\rho/(1+\rho)}\left[ \sum_{b}\left(\sum_{a}Q(a)W(b|a)^{1/(1+\rho)} \right)^{1+\rho}\right]^{1/(1+\rho)} \nonumber \\
&= \Lambda_\rho\left( \frac{\rho}{1+\rho}\right) + \frac{\log \sum_{y}\left[\sum_{x}Q(x)W(y|x)^{1/(1+\rho)} \right]^{1+\rho}}{(1+\rho)} \label{eq:thrm1-pf-step5-13} \\
&=2\Lambda_\rho\left( \frac{\rho}{1+\rho}\right), \label{eq:thrm1-pf-step5-14}
\end{align}
where \eqref{eq:thrm1-pf-step5-13} follows from the definition of $\Lambda_\rho(\cdot)$ and \eqref{eq:thrm1-pf-step5-14} follows from \eqref{eq:lem5-0}. Plugging \eqref{eq:thrm1-pf-step5-14} into \eqref{eq:thrm1-pf-step5-9.1} yields \eqref{eq:thrm1-pf-step5-8.1}, which was to be shown.
\end{itemize}


\section{A Concentration Upper Bound For Sums of I.I.D.\,Random Variables}
\label{app-rv-EA}

Let $\{Z_{n}\}_{n=1}^N$ be i.i.d.\,random variables with law $\nu$. Assume $|Z_{n}| \in \bbR$ $\nu$-(a.s.) and $\textrm{Var}[Z_{n}] >0$. Moreover, let $\Lambda(\lambda) \eqdef \log \mE[e^{\lambda Z_{1}}]$, $\hat{S}_N \eqdef \frac{1}{N}\sum_{n=1}^N Z_{n}$ and $\mu_N$ denote the law of $\hat{S}_N$. 

Consider some $q_N$ and assume there exists $\eta_N>0$ such that 
\begin{itemize}
\item[(i)] There exists a neighborhood of $\eta_N$, such that $\Lambda(\lambda) < \infty$ for all $\lambda$ in this neighborhood. 
\item[(ii)] $\Lambda^\prime(\eta_N) = q_N$. 
\end{itemize}

Observe that owing to the property (i) above, $\Lambda(\cdot)$ is infinitely differentiable at $\eta_N$. 

We aim to derive a sharp upper bound on $\mu_N([q_N,\infty))$. Note that this problem is well-studied in probability theory and indeed $\mu_N([q_N,\infty))$ is asymptotically characterized both for fixed-threshold sets \cite{bahadur-rao60}, and varying-threshold sets \cite{chaganty-sethuraman93}. However, both of these results require the sequence of random variables to be either lattice\footnote{A random variable $T$ is called lattice if there exist constants $c$ and $h \in \bbR^+$ such that $T \in \{ c + k h : k \in \bbZ\}-\textnormal{(a.s.)}$. Here, $c$ (resp. $h$) is called the displacement (resp. span) of the random variable \cite[pg.~129]{durrett05}.} or non-lattice throughout the sequence and the regularity conditions necessary for their validity in case of lattice random variables turns out to be tedious in our application.
Therefore, we prove Lemma~\ref{lem:EA-new} below, which is valid regardless of the lattice nature of the random variables and holds for any $N \in \bbZ^+$, although the constant term is weaker than the result of \cite{chaganty-sethuraman93}. The proof is essentially the same as Dembo-Zeitouni's proof of \cite[Theorem~3.7.4]{dembo-zeitouni98}. The main difference is we use the Berry--Esseen Theorem \cite[Chapter~III]{esseen45}, which is valid regardless of whether the random variables are lattice, instead of the Berry--Esseen expansion \cite[Chapter~IV]{esseen45}, which necessitates one to distinguish between lattice and non-lattice random variables. The proof is included for completeness. 

To state the lemma, we define $\tilde{\nu}_N$ such that 
\begin{equation}
\frac{d \tilde{\nu}_N}{d \nu}(z) \eqdef e^{z\eta_N  - \Lambda(\eta_N)}. 
\label{eq:app-rv-EA-1}
\end{equation}
Further, define $T_{n,N} \eqdef \frac{Z_{n} - \Lambda^\prime(\eta_N)}{\sqrt{\Lambda^{\prime \prime}(\eta_N)}}$, let $\Lambda^\ast(q_N)$ denote the Fenchel-Legendre transform of $\Lambda(\cdot)$ at $q_N$, i.e., 
\begin{equation}
\Lambda^\ast(q_N) \eqdef \sup_{\lambda \in \bbR} \, \{ \lambda q_N - \Lambda_N(\lambda)\},
\label{eq:app-rv-EA-3}
\end{equation}
and $m_{3,N} \eqdef \mE_{\tilde{\nu}_N}[|T_{n,N}|^3]$.

\begin{lemma}For any $N \in \bbZ^+$, 
\begin{equation}
\mu_N([q_N,\infty)) \leq e^{-N \Lambda^\ast(q_N)} \frac{1}{\sqrt{N}}\left\{ \frac{m_{3,N}}{\Lambda^{\prime \prime}(\eta_N)^{3/2}} + \frac{1}{\sqrt{2\pi \Lambda^{\prime \prime}(\eta_N)}\eta_N } \right\}.
\label{eq:app-rv-EA-final}
\end{equation}
\label{lem:EA-new}
\eot
\end{lemma}

\begin{IEEEproof}
First, note that since $Z_{n}$ is real-valued $\nu$-(a.s.), \eqref{eq:app-rv-EA-1} implies that $\nu$ and $\tilde{\nu}_N$ are equivalent probability measures. Also, it is not hard to check that 
\begin{equation}
\mE_{\tilde{\nu}_N}[Z_{n}] = \Lambda^\prime(\eta_N), \, \mV_{\tilde{\nu}_N}[Z_{n}] = \Lambda^{\prime \prime}(\eta_N).
\label{eq:app-rv-EA-2}
\end{equation}
Using \eqref{eq:app-rv-EA-2} and the fact that $\textrm{Var}[Z_{n}] >0$, one can deduce that $\Lambda^{\prime \prime}(\eta_N) >0$. 

Next, define $W_N \eqdef \frac{1}{\sqrt{N}} \sum_{n=1}^N T_{n,N}$. Since $\Lambda^\prime(\eta_N) = q_N$ and $\Lambda^{\prime \prime}(\eta_N) >0$, it is easy to see that $\eta_N$ is the unique maximizer of the right side of \eqref{eq:app-rv-EA-3}. 

One can check that 
\begin{equation}
\mu_N([q_N,\infty)) = e^{-N \Lambda^\ast(q_N)} \int_{0}^{\infty} e^{-x \eta_N \sqrt{N \Lambda^{\prime \prime}(\eta_N)} }dF_N(x), 
\label{eq:app-rv-EA-4}
\end{equation}
where $F_N$ is the distribution of $W_N$ when $Z_{n}$ are i.i.d.\,with $\tilde{\nu}_N$. By using integration by parts, along with elementary calculation, one can verify that  
\begin{equation}
\int_{0}^{\infty} e^{-x \eta_N \sqrt{N\Lambda^{\prime \prime}(\eta_N)} }dF_N(x) = \int_{0}^\infty e^{-t}\left[ F_N\left( \frac{t}{\eta_N \sqrt{N\Lambda^{\prime \prime}(\eta_N)}}\right) - F_N(0)\right] dt. 
\label{eq:app-rv-EA-5}
\end{equation} 

An application of the Berry-Esseen theorem (e.g.,~\cite[eq.~(III.$15^\prime$)]{esseen45}) yields\footnote{For the sake of notational convenience, we take the universal constant in the theorem as $1/2$, although it is not the best known constant for the case of i.i.d. random variables. See \cite{korolev2010} for a recent survey of the best known constants in the Berry-Esseen theorem.}
\begin{equation}
F_N\left( \frac{t}{\eta_N \sqrt{N \Lambda^{\prime \prime}(\eta_N)}}\right) - F_N(0) \leq \Phi\left( \frac{t}{\eta_N \sqrt{N\Lambda^{\prime \prime}(\eta_N)}}\right) - \Phi(0) + \frac{ m_{3,N}}{\Lambda^{\prime \prime}(\eta_N)^{3/2}}\frac{1}{\sqrt{N}}. \label{eq:app-rv-EA-6}
\end{equation}
Via a power series expansion around $0$ and using the fact that $\phi^\prime(\cdot) \leq 0$ on $\bbR_+$, we deduce that  
\begin{equation}
\Phi\left( \frac{t}{\eta_N \sqrt{N\Lambda^{\prime \prime}(\eta_N)}}\right) - \Phi(0)  \leq \frac{t }{\eta_N \sqrt{2\pi N\Lambda^{\prime \prime}(\eta_N)}}. 
\label{eq:app-rv-EA-7}
\end{equation}
Plugging \eqref{eq:app-rv-EA-6} and \eqref{eq:app-rv-EA-7} into the right side of \eqref{eq:app-rv-EA-5} and carrying out the  integration, we have
\begin{equation}
\int_{0}^{\infty} e^{-x\eta_N \sqrt{N \Lambda^{\prime \prime}(\eta_N)} }dF_N(x) \leq \frac{1}{\sqrt{N}}\left\{ \frac{m_{3,N}}{\Lambda^{\prime \prime}(\eta_N)^{3/2}} + \frac{1}{\eta_N\sqrt{2\pi \Lambda^{\prime \prime}(\eta_N)} } \right\}. \label{eq:app-rv-EA-final-0}
\end{equation}
Plugging \eqref{eq:app-rv-EA-final-0} into \eqref{eq:app-rv-EA-4} yields \eqref{eq:app-rv-EA-final}. 
\end{IEEEproof}


\section{Proof of Lemma~\ref{lem:lem-step5}}
\label{app-lem-step5}

We first claim that 
\begin{equation}
\mV_{\tilde{Q}_{X,Y,Z}^{\tilde{\bv}, \rho}}\left[ \log \frac{W(Y|X)}{f_\rho}\right], \mV_{\tilde{Q}_{X,Y,Z}^{\tilde{\bv}, \rho}}\left[ \log \frac{W(Y|Z)}{W(Y|X)}\right] \in \bbR^+. 
\label{eq:thrm1-pf-step5-8}
\end{equation}

To see \eqref{eq:thrm1-pf-step5-8}, note that 
\begin{align}
\left[\mV_{\tilde{Q}_{X,Y,Z}^{\tilde{\bv}, \rho}}\left[ \log \frac{W(Y|X)}{f_\rho(Y)}\right] =0 \right] & \Longleftrightarrow \left[ \log \frac{W(y|x)}{f_\rho(y)} = - \Lambda_\rho^\prime\left( \frac{\rho}{1+\rho}\right), \, \forall (x,y) \in \cS_Q \right] \nonumber \\
& \Longrightarrow \left[ \textnormal{the pair }(Q,W) \textrm{ is singular}\right]. \label{eq:thrm1-pf-step5-9}
\end{align}
The right side of \eqref{eq:thrm1-pf-step5-9} yields a contradiction, hence we conclude that $\mV_{\tilde{Q}_{X,Y,Z}^{\tilde{\bv}, \rho}}\left[ \log \frac{W(Y|X)}{f_\rho(Y)}\right]  > 0$. 

Similarly, 
\begin{align}
\left[\mV_{\tilde{Q}_{X,Y,Z}^{\tilde{\bv}, \rho}}\left[ \log \frac{W(Y|Z)}{W(Y|X)}\right] =0 \right] & \Longleftrightarrow \left[ \log \frac{W(y|z)}{W(y|x)} = 0, \, \forall (x,y,z) \in \tilde{\cS}_Q \right] \nonumber \\
& \Longrightarrow \left[ \textnormal{the pair }(Q,W) \textrm{ is singular}\right]. \label{eq:thrm1-pf-step5-10}
\end{align}
The right side of \eqref{eq:thrm1-pf-step5-10} yields a contradiction, hence we conclude that $\mV_{\tilde{Q}_{X,Y,Z}^{\tilde{\bv}, \rho}}\left[ \log \frac{W(Y|Z)}{W(Y|X)}\right]  > 0$. 

Further, as an immediate consequence of the nonsingularity of the pair $(Q,W)$, there is no $\alpha \in \bbR$ satisfying 
\[
\log\frac{W(y|z)}{W(y|x)} = \alpha\left( \log\frac{W(y|x)}{f_\rho(y)} + \Lambda_\rho^\prime\left( \frac{\rho}{1+\rho}\right)\right), \, \forall (x,y,z) \in \tilde{\cS}_Q. 
\]

This last observation, coupled with  \eqref{eq:thrm1-pf-step5-8} and the Cauchy-Schwarz inequality, implies \eqref{eq:tilted-cov}, which was to be shown. 


\section{Proof of Lemma~\ref{lem:vector-EA}}
\label{app-vector-EA}
The proof follows from essentially the same arguments as in one dimensional case given in Appendix~\ref{app-rv-EA}. The only significant difference is  the usage  of a ``concentration function'' theorem for sums of independent random vectors by Esseen \cite[Theorem~6.2]{esseen68}, instead of the Berry-Esseen theorem. 

For notational convenience, we define 
\[
\bA_{n}(N) \eqdef \left[ \log \frac{W(Y_n|X_n)}{f_N^\ast(Y_n)}, \, \log\frac{W(Y_n|Z_n)}{W(Y_n|X_n)}\right]^T, \quad\mathbf{S}_N \eqdef \frac{1}{N}\sum_{n=1}^N \bA_n(N), 
\]
and let $\mu_N$ denote the law of $\mathbf{S}_N$ when $\bA_n(N)$ is distributed according to $\tilde{P}_{X,Y,Z}$. Clearly, $\tilde{\alpha}_N = \mu_N(\cB(N))$. 

Define $\bT_n(N) \eqdef \bA_n(N) - \bb(N)$ and $\bW_N \eqdef \frac{1}{\sqrt{N}}\sum_{n=1}^N \bT_n(N)$. Note that 

\begin{align}
\mE_{\tilde{Q}_{X,Y,Z}^{\bv^\ast(N)}}\left[  \left[ \log\frac{W(Y|X)}{f_N^\ast(Y)}, \log \frac{W(Y|Z)}{W(Y|X)}\right]^T  \right] & = \left[ \left.\frac{\d \Lambda_{1, N}(\bv_1, \bv^\ast_2(N))}{\d \bv_1}\right|_{\bv_1 = \bv^\ast_1(N)}, \left.\frac{\d \Lambda_{1, N}(\bv^\ast_1(N), \bv_2)}{\d \bv_2}\right|_{\bv_2 = \bv^\ast_2(N)}\right]^T \label{eq:tilted-mean1} \\
 & =  [-\Lambda_N^\prime(\rho^\ast_N/(1+\rho^\ast_N)),0]^T, \label{eq:tilted-mean2} 
\end{align}
where \eqref{eq:tilted-mean1} follows by evaluating the right sides of \eqref{eq:thrm1-pf-step5-2} and \eqref{eq:thrm1-pf-step5-2.5} in Appendix~\ref{app-lem-diff-new-pf} at $\bv^\ast(N)$ and \eqref{eq:tilted-mean2} follows from item (i) of Lemma~\ref{lem:diff-new}. Equation \eqref{eq:tilted-mean2} ensures that $\textnormal{E}_{\tilde{Q}_{X,Y,Z}^{\bv^\ast(N)}}[\bT_n(N)] = \mathbf{0}$. 

By elementary calculation, one can check that 
\begin{equation}
\mu_N(\cB(N)) = e^{-N \Lambda_{1,N}^\ast(\bb(N))} \int_{0}^\infty  \int_{0}^\infty e^{-\sqrt{N}\langle \bv^\ast(N), \bx \rangle}dF_N(\bx), 
\label{eq:vector-EA-4}
\end{equation}
where $F_N$ is the distribution of $\bW_N$ when $\bA_n(N)$ are i.i.d.\,with $\tilde{Q}_{X,Y,Z}^{\bv^\ast(N)}$. 

Since $e^{-\sqrt{N}\langle \bv, \bx \rangle}$ is a continuous function of bounded variation and $F_N(\bx)$ is a function of bounded variation, we apply the integration by parts formula of Young \cite[Eq. 4]{young18} to deduce that 
\begin{align}
\int_{0}^\infty  \int_{0}^\infty e^{-\sqrt{N}\langle \bv^\ast(N), \bx \rangle}dF_N(\bx) & = \int_{0}^\infty \int_{0}^\infty e^{-\langle \mathbf{1}, \mathbf{t} \rangle} \left[ F_N\left( \frac{t_1}{\bv^\ast_1(N) \sqrt{N}}, \frac{t_2}{\bv_2^\ast(N) \sqrt{N}}\right) - F_N\left(0, \frac{t_2}{\bv^\ast_2(N) \sqrt{N}}\right) \right. \nonumber \\
&  \quad \left. - F_N\left( \frac{t_1}{\bv^\ast_1(N) \sqrt{N}}, 0\right) + F_n\left( 0, 0\right) \right] dt_1 dt_2 \nonumber \\
& =  \int_{0}^\infty \int_{0}^\infty e^{-\langle \mathbf{1}, \mathbf{t} \rangle} \Pr \left\{ \mathbf{W}_N \in \left(0, \frac{t_1}{\bv^\ast_1(N) \sqrt{N}} \right] \times \left(0, \frac{t_2}{\bv^\ast_2(N) \sqrt{N}} \right]\right\} dt_1 dt_2, \label{eq:vector-EA-6}
\end{align} 
where the probability is computed when $A_n(N)$ are i.i.d.\,with $\tilde{Q}_{X,Y,Z}^{\bv^\ast(N)}$. 

In order to conclude the proof, we upper bound the right side of \eqref{eq:vector-EA-6} by using a concentration inequality of Esseen \cite[Corollary of Theorem~6.2]{esseen68}. To state his result, we need the following definitions.

Let $\bT_n^s(N) \eqdef \bT_n(N) - \bT_n^\prime(N)$, where $\bT_n^\prime(N)$ and $\bT_n(N)$ are i.i.d. Let $\tilde{\nu}^s_N$ denote the law of $\bT_n^s(N)$. Following \cite[eq.~(6.4)]{esseen68}, define
\begin{equation*}
\kappa_N(u) \eqdef \inf_{|\bt| =1}\int_{|\bx| < u} (\langle \bt, \bx \rangle)^2 d\tilde{\nu}_N^s(\mathbf{x}). 
\end{equation*}
Finally, let $\mathcal{S}_{\rho}(\mathbf{c}_\mo)$ denote the sphere in $\bbR^2$ with radius $\rho$ and center $\mathbf{c}_\mo$. 

In our case, \cite[Corollary to Theorem~6.2]{esseen68} reads as follows: for any $\rho \in \bbR^+$, 
\begin{equation}
\sup_{\bc_\mo \in \bbR^2}\Pr\left\{ \sum_{n=1}^N \mathbf{A}_n(N) \in \mathcal{S}_{\rho}(\mathbf{c}_\mo) \right\} \leq c \left( \frac{\rho}{\tau} \right)^2 \left( \frac{1}{N \sup_{u \geq \tau} u^{-2}\kappa_N(u)} \right), \, \forall \, \tau \in (0, \rho],
\label{eq:vector-EA-6.1}
\end{equation}
where $c$ is a universal constant that only depends on the dimension of the random vector, which is $2$ in our case. 

Next, we explain how to use \eqref{eq:vector-EA-6.1} to conclude the proof. Since 
\[
\lim_{N \rightarrow \infty}\bA_n(N) = \bA_n \eqdef \left[ \log \frac{W(Y_n|X_n)}{f^\ast(Y_n)}, \, \log\frac{W(Y_n|Z_n)}{W(Y_n|X_n)}\right]^T, \, \tilde{P}_{X,Y,Z}-\textrm{(a.s.)},
\]
$\bA_n$ is bounded almost surely under $\tilde{P}_{X,Y,Z}$. Further, $\tilde{Q}_{X,Y,Z}^{\bv^\ast(N)}$ is equivalent to $\tilde{P}_{X,Y,Z}$ for all $N$. These two observations imply that there exists $k(R,W,Q) \in \bbR^+$ and a sufficiently large $N_1$ that only depends on $R$, $W$ and $Q$ such that $\max\{\bT_{1,n}(N)^s, \bT_{2,n}(N)^s\} \leq k(R,W,Q)$, almost surely under $\tilde{\nu}_N^s$ for all $N \geq N_1$. 

Consider any $N \geq N_1$ from now on. One can also check that
\begin{equation*}
\bS_N^s \eqdef \mE_{\tilde{\nu}^s}\left[ \bT_n^s(N) \bT_n^s(N)^T \right] = 2 \bS_N, 
\label{eq:vector-EA-8}
\end{equation*}  
which, in turn, implies that for any $u \geq k(R,W,Q)$,
\begin{equation}
\kappa_N(u) =  \inf_{|\bt|=1}\int (\langle \bt, \bx \rangle)^2 d\tilde{\nu}_N^s(\mathbf{x}) = \inf_{|\bt|=1} \bt^T \bS_N^s \bt = 2 \inf_{|\bt|=1} \bt^T \bS_N \bt = 2 \lambda_{\textrm{min}}(\bS_N). \label{eq:vector-EA-9}
\end{equation}
Since $\textrm{det}(\bS_N) > 0$, which follows from Lemma~\ref{lem:lem-step5}, we also have $\lambda_{\textrm{min}}(\bS_N) >0$. 

By letting $\rho \eqdef \sqrt{\frac{t_1^2}{(\bv_1^\ast(N))^2} + \frac{t_2^2}{(\bv_2^\ast(N))^2}}$, $\mathbf{c}_\mo = \mathbf{0}$ and $\tau = \rho$, \eqref{eq:vector-EA-6.1} implies that  
\begin{align}
\Pr \left\{ \mathbf{W}_N \in \left(0, \frac{t_1}{\bv^\ast_1(N) \sqrt{N}} \right] \times \left(0, \frac{t_2}{\bv^\ast_2(N) \sqrt{N}} \right]\right\} & \leq \Pr\left\{ \sum_{n=1}^N \mathbf{A}_n(N) \in \cS_{\rho}(\mathbf{c}_\mo) \right\} \nonumber \\
 & \leq \frac{c}{N} \inf_{u \geq \rho} \frac{u^2}{\kappa_N(u)} \nonumber \\
 & \leq \frac{c}{2\lambda_{\min}(\mathbf{\Sigma}_N)N}\left( k(R,W,Q)^2 + \frac{t_1^2}{(\bv_1^\ast(N))^2}+ \frac{t_2^2}{(\bv_2^\ast(N))^2}\right), \label{eq:vector-EA-10}
\end{align}
where \eqref{eq:vector-EA-10} follows from \eqref{eq:vector-EA-9}. By substituting \eqref{eq:vector-EA-10} into \eqref{eq:vector-EA-6} and carrying out the calculation, we deduce that 

\begin{equation*}
\int_{0}^\infty \int_{0}^\infty e^{-\sqrt{N}\langle \mathbf{v}^\ast(N), \mathbf{x} \rangle} dF_N(\mathbf{x}) \leq \frac{c}{2\lambda_{\min}(\mathbf{\Sigma}_N)N}\left( k(R,W,Q)^2 + \frac{2}{(\bv_1^\ast(N))^2} + \frac{2}{(\bv_2^\ast(N))^2}\right), 
\end{equation*}
which, in light of \eqref{eq:vector-EA-4}, suffices to conclude the proof. 

\section*{acknowledgment} 
The authors would like to thank Alfred O.~Hero~III for helpful discussions surrounding Definition~\ref{def:singular}. The first author thanks Paul Cuff for his hospitality while portions of this work were being completed during his visit to Princeton University. This research was supported by the National Science Foundation under grant CCF-1218578.

\end{document}